\documentclass{aa}
\usepackage{epsfig}
\usepackage{txfonts}
\def\d{\displaystyle}
\def\kms{km\,s$^{-1}$}

\def\vs{$v\sin{i}$}
\def\te{$T_{\rm eff}$}
\def\lg{$\log{g}$}
\def\ha{H$_{\alpha}$}
\def\hb{H$_{\beta}$}
\def\ebv{$E(B$$-$$V)$}

\begin{document}

\title{Spectral analysis of Kepler SPB and $\beta$\,Cep candidate stars\thanks{Based 
on observations with the 2-m Alfred Jensch telescope at the Th\"uringer
Landessternwarte (TLS) Tautenburg.}}
\author{H. Lehmann\inst{1} 
\and A. Tkachenko\inst{1} 
\and T. Semaan\inst{2}
\and J. Guti\'{e}rrez-Soto\inst{2,3}
\and B. Smalley\inst{4}
\and M. Briquet\inst{5}
\and D. Shulyak\inst{6}
\and V. Tsymbal\inst{7}
\and P. de Cat\inst{8}
}
\institute{ 
Th\"uringer Landessternwarte Tautenburg, 07778 Tautenburg, Germany\\
\email{lehm@tls-tautenburg.de, andrew@tls-tautenburg.de} \and
GEPI, Observatoire de Paris, CNRS, Universit\'{e} Paris Diderot, 5 place
Jules Janssen, 92190 Meudon, France\\
\email{thierry.semaan@obspm.fr}
\and
Instituto de Astrof\'{i}sica de Andaluc\'{i}a (CSIC), Apartado 3004, 18080
Granada, Spain,  
\email{jgs@iaa.es}
\and
Astrophysics Group, Keele University, Staffordshire, ST5 5BG, United Kingdom,
\email{bs@astro.keele.ac.uk}
\and
Instituut voor Sterrenkunde, Katholieke Universiteit Leuven, Belgium,
\email{maryline@ster.kuleuven.be}
\and
Georg-August-University, G\"ottingen, Germany,
\email{denis.shulyak@gmail.com} \and
Tavrian National University, Department of Astronomy, Simferopol, Ukraine,
\email{vadim@starsp.org}
\and
Royal Observatory of Belgium,
\email{peter@oma.be} }

\date{Received ; accepted}

\abstract  
{For an asteroseismic modeling, the analysis of the high-accuracy light curves delivered by the Kepler satellite
mission needs support by ground-based multi-colour and spectroscopic observations.}
{We determine the fundamental parameters of SPB and $\beta$\,Cep candidate stars observed by the
Kepler satellite mission and estimate the expected types of non-radial pulsators.}
{We compare newly obtained high-resolution spectra with synthetic spectra computed on a grid of stellar parameters 
assuming LTE
and check for NLTE effects for the hottest stars. For comparison, we determine \te\ independently from fitting
the spectral energy distribution of the stars obtained from the available photometry.}
{We determine \te, \lg, micro-turbulent velocity, \vs, metallicity, and elemental abundance for 14 of the 16 
candidate stars, two of the stars are spectroscopic binaries.
No significant influence of NLTE effects on the results could be found. 
For hot stars, we find systematic deviations of the determined effective temperatures from those given in the
Kepler Input Catalogue. The deviations are confirmed by the results obtained from ground-based photometry. 
Five stars show reduced metallicity, two stars are He-strong, one is He-weak,
and one is Si-strong. Two of the stars could be $\beta$\,Cep/SPB hybrid pulsators, four
SPB pulsators, and five more stars are located close to the borders of the SPB instability region.}
{}


\maketitle

\section{Introduction}
The Kepler satellite delivers light curves of unique accuracy and time coverage, 
providing unprecedented data for the asteroseismic analysis. The identification of 
non-radial pulsation modes and the asteroseismic modeling, on the other hand, 
require a classification of the observed stars in terms of \te, 
\lg, \vs, and metallicity. These basic stellar parameters can only be 
obtained from the colors of the stars which are not measured by the satellite. 
That is why ground-based multi-color and spectroscopic observations of the 
Kepler target stars are urgently needed. We describe a semi-automatic method 
of spectrum analysis based on high-resolution spectra of stars in the Kepler 
satellite field of view that have been proposed 
by the Working Groups 3 and 6 of the Kepler Asteroseismic Science Consortium (KASC)
to be candidates for SPB and $\beta$\,Cep pulsators.

The object selection was mainly based on the data given in Kepler Input Catalogue (KIC).
The spectral types given for these stars in the SIMBAD database are based on only a few, 
older measurements and 
in most cases no luminosity class is given. For the KIC data, on the other hand, an uncertainty
of the surface gravity of $\pm$0.5 dex is stated in the catalogue, much too high for an accurate classification in terms
of non-radial pulsators, and there are hints that the temperature values given in the KIC 
show larger deviations for the hotter stars (see Molenda-\.Zakowicz et al. \cite{Molenda}).
For two of the selected stars we found no classification 
in the SIMBAD database and for two no classification in the KIC.

The aim of this work is to provide fundamental stellar parameters like effective temperature \te, 
surface gravity \lg, metallicity $\epsilon$, and projected rotation velocity \vs\ for an asteroseismic
modeling of the stars, to compare our results of spectral analysis with the KIC data 
and to estimate the expected type of variability of the different target stars.

For the analysis, we used stellar atmosphere models and synthetic spectra computed under 
the LTE assumption. The advantage of the applied programs (see Sect.\,3) is that they allow to 
use different metallicity and individual abundances of He and metals and that they are
running fast in parallel mode on a cluster PC installed at TLS. 

In the investigated spectral region, the spectra of the B-type stars are dominated by the \ion{He}{I}
lines. Auer \& Mihalas (\cite{Auer}) state that we have to expect deviations in
the equivalent widths of the He lines due to the effects of departure from local thermodynamic 
equilibrium (LTE) for stars 
in the 15\,000 to 27\,500 K temperature range in the order of 10\%  for 
4026\,\AA$<$$\lambda$$<$5047\AA\ and of up to 30\% for the 5\,876\,\AA\ line.
There is a controversial discussion about these results, however (see Sect.\,6.1). 
For that reason, to check for the reliability of our results of the spectral analysis
obtained in the LTE regime, we repeated the analysis for the four hottest stars of our sample 
using programs with non-local thermodynamic equilibrium (NLTE) capability (Sect.\,4) but had 
to assume constant, solar metallicity and He abundance. 

One further test is described in Sect.\,5, where we derive the effective temperatures
of the stars from their available photometry and try to explain the deviations of the \te\
given in the KIC from our spectroscopically determined values. The results obtained from 
the different methods are discussed in Sect.\,6. 

\section{Observations and spectrum reduction}
Spectra of 16 bright (V$<$10.4) suspected SPB and $\beta$ Cep stars selected from the KIC 
have been taken with the Coude-Echelle spectrograph attached to the 2-m 
telescope at the Th\"uringer Landessternwarte (TLS) Tautenburg. 
The spectra have a resolution of 32\,000 and cover the 
wavelength range from 470 to 740 nm. Table\,\ref{objects} gives KIC number, common designation,
suspected type of variability, $V$-magnitude, number of observed spectra, and the signal-to-noise of
the averaged spectra for all observed stars. Spectra have been reduced using standard ESO-MIDAS
packages. The reduction included filtering of cosmic rays, bias and stray light subtraction,
flat fielding, optimum extraction of the Echelle orders, wavelength calibration using a
Th-Ar lamp, normalization to the local continuum, and merging of the orders. Small shifts of
the instrumental zero-point have been corrected by an additional wavelength calibration using
a large number of telluric O$_2$-lines. 
Finally, the difference in the radial velocities (RVs) between all spectra of the same star have been determined
from cross-correlation and the spectra have been rebinned according to theses differences
and added to build the mean, averaged spectrum. 

\begin{table}
\tabcolsep 1.5mm
\caption{List of observed stars.}
\begin{tabular}{rllrcr}
\hline\hline
\multicolumn{1}{c}{KIC} & designation & type & \multicolumn{1}{c}{$V$} & $N$ & S/N \\
\hline
 3\,240\,411 & GSC 03135$-$00115 & $\beta$\,Cep      & 10.2 & 2 &  67 \\
 3\,756\,031 & GSC 03135$-$00619 & $\beta$\,Cep      & 10.0 & 2 &  80 \\
 5\,130\,305 & HD ~226\,700	 & SPB  	     & 10.2 & 2 &  74 \\
 5\,217\,845 & HD ~226\,628	 & SPB  	     &  9.3 & 2 & 103 \\
 5\,479\,821 & HD ~226\,795	 & SPB  	     &  9.9 & 1 &  65 \\
 7\,599\,132 & HD ~180\,757	 & SPB  	     &  9.3 & 1 &  75 \\
 8\,177\,087 & HD ~186\,428	 & SPB, $\beta$\,Cep &  8.1 & 1 & 138 \\
 8\,389\,948 & HD ~189\,159	 & SPB  	     &  9.1 & 1 &  81 \\
 8\,451\,410 & HD ~188\,459	 & SPB  	     &  9.1 & 2 & 104 \\
 8\,459\,899 & HD ~190\,254	 & $\beta$\,Cep      &  8.7 & 2 & 127 \\
 8\,583\,770 & HD ~189\,177	 & SPB  	     & 10.1 & 2 &  74 \\
 8\,766\,405 & HD ~187\,035	 & SPB, $\beta$\,Cep &  8.8 & 1 & 103 \\
10\,960\,750 & BD+482\,781	 & SPB, $\beta$\,Cep &  9.7 & 1 &  59 \\
11\,973\,705 & HD ~234\,999	 & SPB  	     &  9.1 & 2 & 116 \\
12\,207\,099 & BD+502\,787	 & $\beta$\,Cep      & 10.3 & 2 &  69 \\
12\,258\,330 & HD ~234\,893	 & SPB, $\beta$\,Cep &  9.3 & 2 & 105 \\
\hline
\end{tabular}\label{objects}
\end{table}

\section{LTE based analysis}
\subsection{The method}

Due to the large \vs\ of many of the observed stars it is impossible to find
a sufficiently large number of unblended spectral lines for a spectral analysis
based on the comparison of the equivalent widths of the lines of single elements.
Thus we decided to analyse the spectra by computing synthetic spectra for a
wider spectral region, including H$_{\beta}$ and a larger number of metal lines. We used
the range from 472 to 588 nm (lower end of the covered wavelength range up to 
the wavelength where stronger telluric lines occur). 

We used the LLmodels program 
(Shulyak et al. \cite{Shulyak}) in its most recent parallel version to compute the atmosphere 
models and the SynthV program (Tsymbal 1996) in a parallelized version written 
by A.T. to compute the synthetic spectra. 
The LLmodels code is a 1-D stellar model atmosphere 
code for early and intermediate type stars assuming LTE which is intended for as accurate a treatment as 
possible of the line opacity using a direct method for the line blanketing calculation.
This line-by-line method is free of any approximations so that it fully describes the dependence 
of the line absorption coefficient on frequencies and depths in a model atmosphere, 
it does not require pre-calculated opacity tables.
The code is based on modified ATLAS9 subroutines (Kurucz \cite{KuruczA}) and
the continuum opacity sources and partition functions of iron-peak elements from ATLAS12 
(Kurucz \cite{KuruczB}) are used.
Like the SynthV program, it can handle individual elemental abundances. Actually, the 
main limitation with respect to hot stars is that both programs assume LTE. 

The line tables have been taken from the VALD data base (Kupka et al. \cite{Kupka}).
They are adjusted by the mentioned programs according to the different spectral types.
For the comparison with the synthetic spectra, the
observed spectra have been rebinned in wavelength according to
their RVs obtained from the cross-correlation with the computed spectra and averaged in
the case of several observations per star. 

We computed the synthetic spectra on a grid in $T_{\rm eff}$, $\log{g}$, $v\sin{i}$, 
and metallicity $\epsilon$, based on a pre-calculated library of atmosphere models. 
The models have been computed by scaling the solar metal abundances
according to the different metallicities from $-$0.8 to +0.8 dex, assuming constant, solar He
abundance and a micro-turbulent velocity of 2\,\kms. At this point, we derived the value of
the metallicity and its error that we will give later in Table\,\ref{ResPar}.

In a second step, we fixed the previously derived parameters to its optimum values and
took the abundance table corresponding to the determined metallicity as the starting point
to readjust the abundances of He and all metals for which we found measurable contributions
in the observed spectra. Here, we iterated the individual abundances together with \vs.

\begin{figure*}\centering
\epsfig{figure=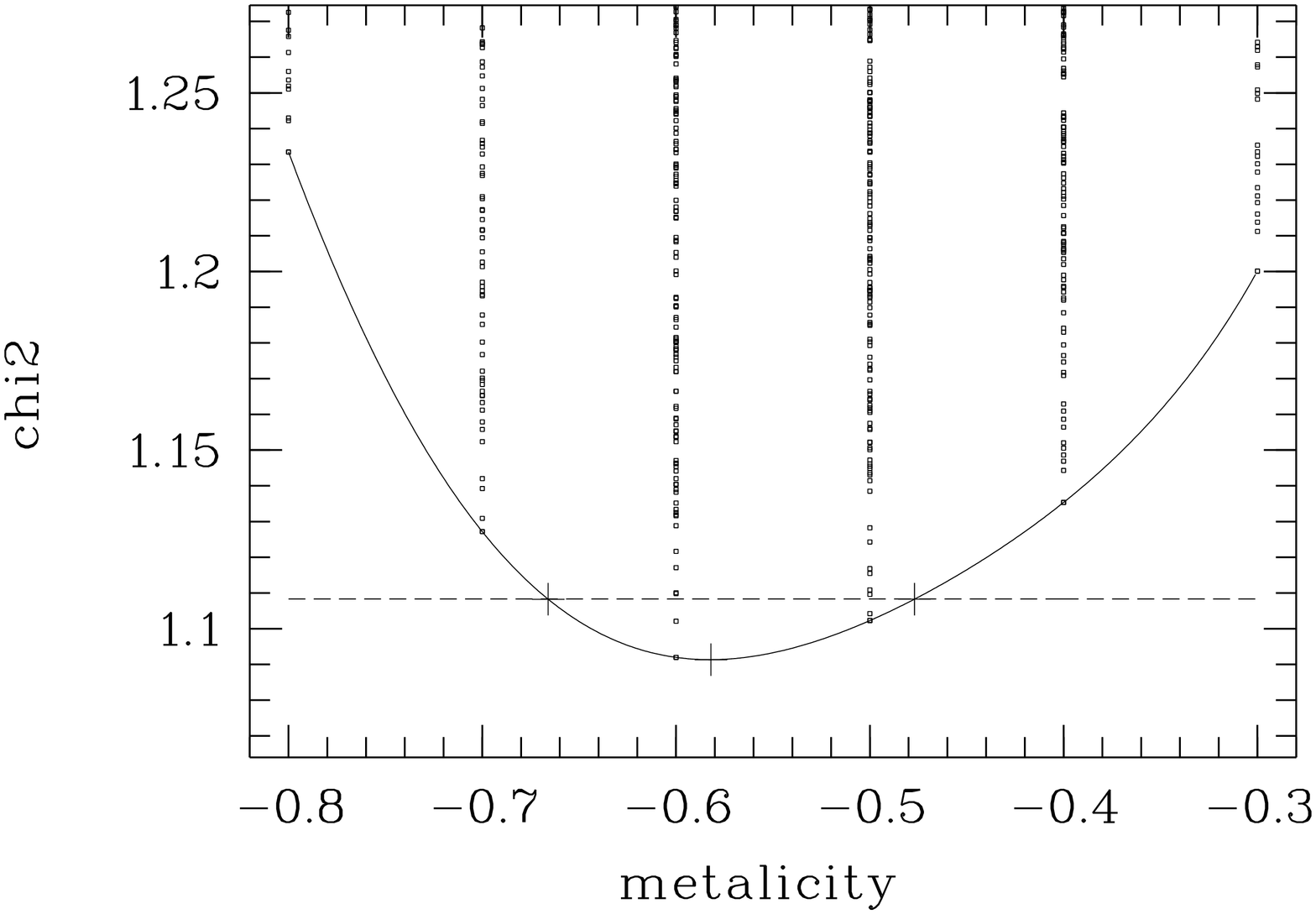,  angle=-90, width=6cm, clip=}
\epsfig{figure=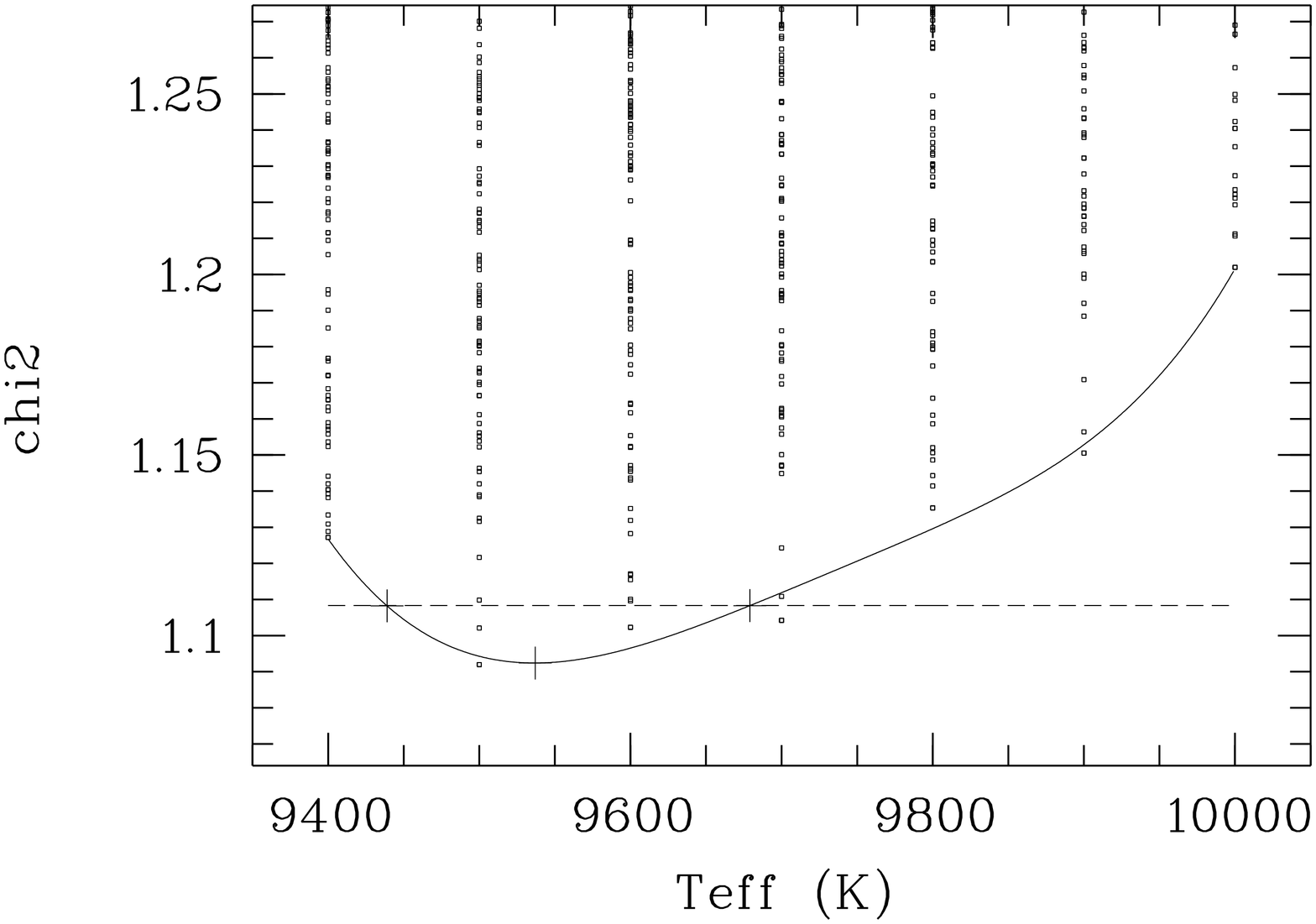,   angle=-90, width=6cm, clip=}
\epsfig{figure=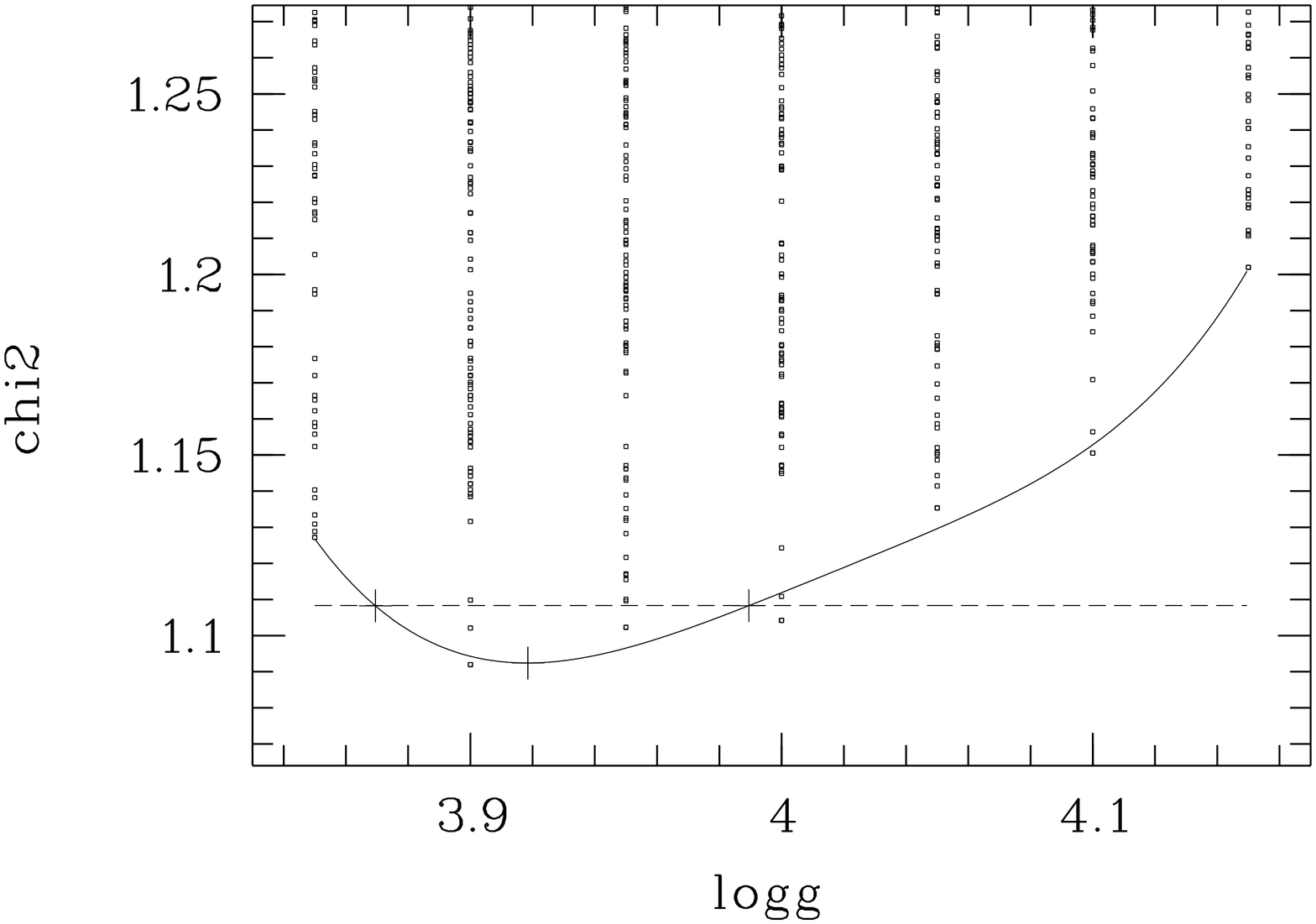,   angle=-90, width=6cm, clip=}\\
\epsfig{figure=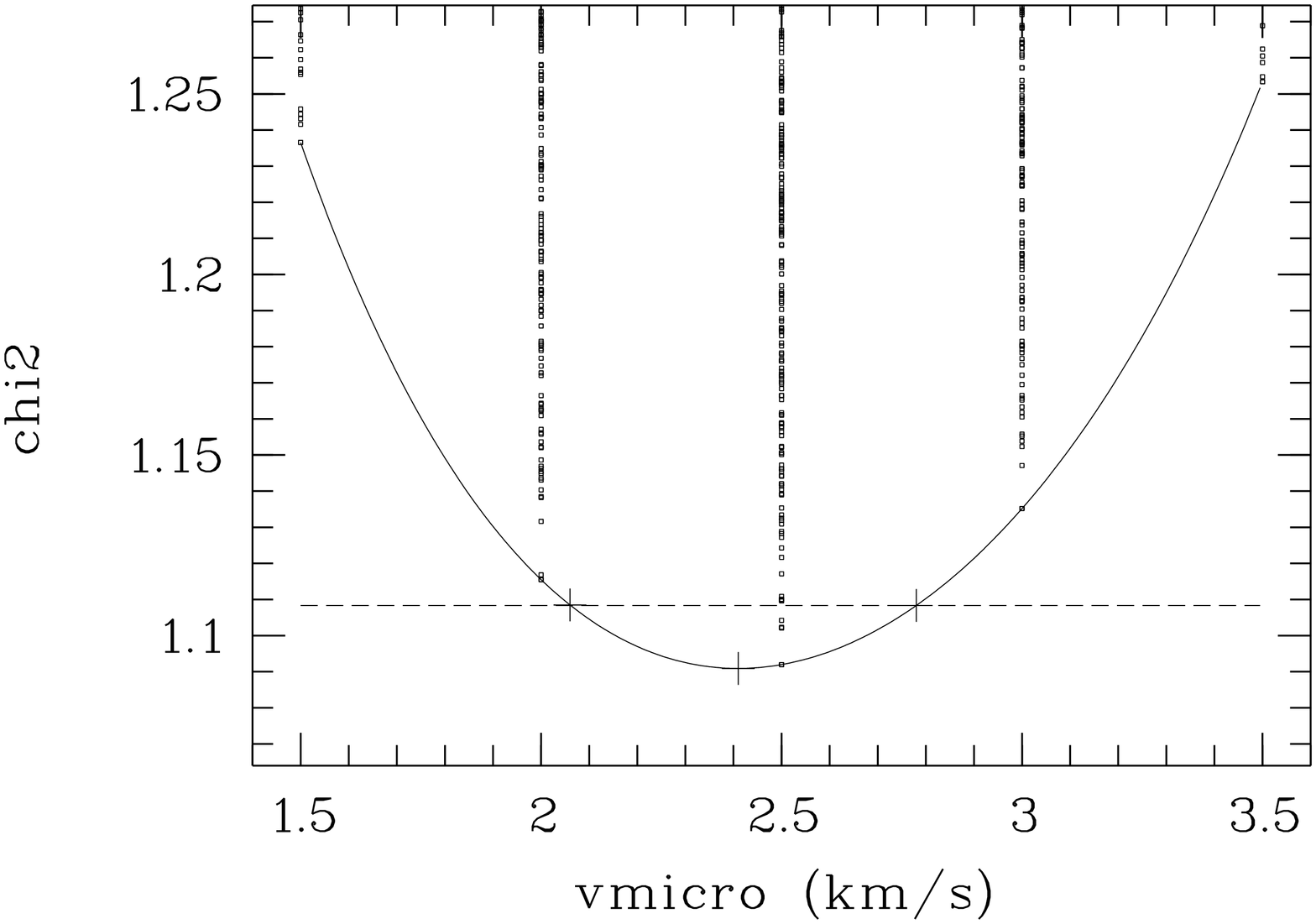, angle=-90, width=6cm, clip=}
\epsfig{figure=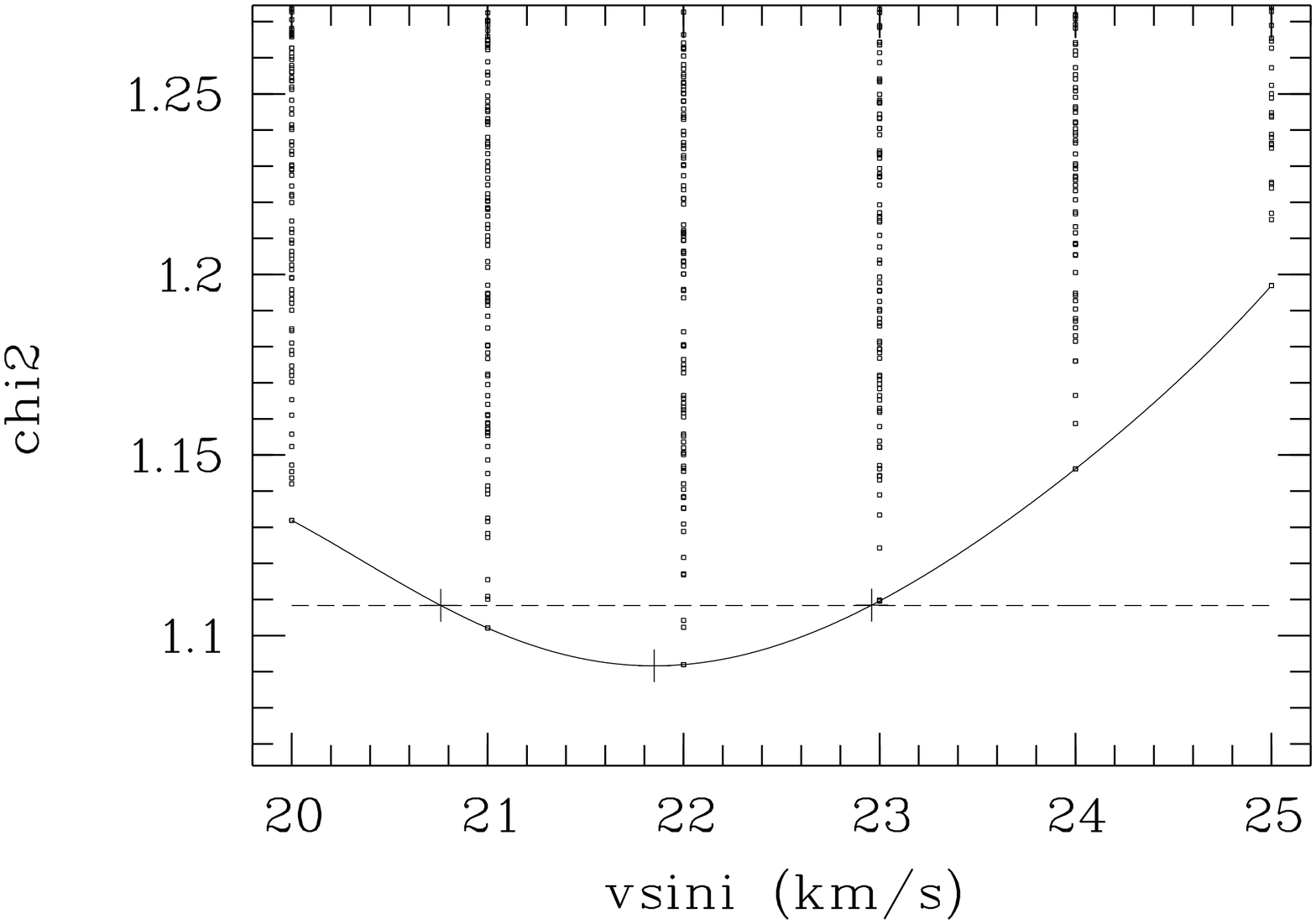,  angle=-90, width=6cm, clip=}
\caption{$\chi^2$ values obtained from the 5-dimensional grid versus the 5 different parameters.} 
\label{VegaFig}
\end{figure*}

In the final step, we readjusted the values of \te, \lg, and \vs\ based on the abundances determined in step two
and added the micro-turbulent velocity $\xi_t$ as a free parameter.
We were not able to compute all the atmosphere models for the full parameter space including the 
individual abundances of He and metal lines and different $\xi_t$, however. We 
used the atmosphere models computed for the fixed, optimum metallicity determined in step 1
but computed the synthetic spectra with the SynthV program based on the individual abundances 
determined in step 2. A comparison of the derived metallicities and metal abundances showed that
the derived values are compatible in most cases.
Only for the three stars where we
found larger deviations of the He abundance (the He weak star KIC\,5\,479\,821 and the He strong stars 
KIC\,8\,177\,087 and KIC\,12\,258\,330), we had to calculate and use atmosphere models based on exactly the determined
He abundances. The final values of the parameters and its errors have been taken from step 1 for the metallicity,
from step 2 for the individual abundances (here, we did no error calculation but assume the error derived for
the metallicity to be the typical error), and from the last step for all the other parameters.

The applied method of grid search allows to find the global minimum of $\chi^2$ and for a realistic estimation 
of the errors of the parameters as we will show in the next section.

\subsection{Testing the method}
The method has been tested on a spectrum of Vega taken with the same instrument
and resolution, with the aim to check for the reliability of the obtained 
values and for the influence of the different parameters on the accuracy of the results. 
Fig.\,\ref{VegaFig}  shows the $\chi^2$-distributions obtained from the grid search. 
Each panel contains all $\chi^2$ values up to a certain value
obtained from all combinations of the different parameters versus one of the parameters.
The dashed lines indicate the 1$\sigma$ confidence
level obtained from the $\chi^2$-statistics assuming that for a large number of 
degrees of freedom, the $\chi^2$-distribution
approaches a Gaussian one. The continuous curves show the polynomial fit to the smallest
$\chi^2$-values. The three crosses in each panel
show the optimum value and the $\pm 1\sigma$ error limits of the corresponding parameter.

Table\,\ref{VegaRes} lists the resulting values. Table\,\ref{VegaComp} compares 
the results with values from the literature. Our values of
$T_{\rm eff}$ and $\log{g}$ agree well with those from previous investigations.
The metallicity and micro-turbulent velocity have been used by us as free parameters 
as well. The obtained values confirm those assumed by the other authors and the 
obtained value of \vs\ is identical with that measured by Hill et al. (\cite{Hill}). 

\begin{table}
\tabcolsep 1.2mm
\renewcommand{\arraystretch}{1.7}
\caption{Fundamental parameters of Vega obtained from grid search.}
\begin{tabular}{lll}
\hline\hline
$T_{\rm eff} = 9540^{+140}_{-100}$ K & 
$\xi_t = 2.41^{+0.37}_{-0.35}$ km\,s$^{-1}$&
$\log(g) = 3.92^{+0.07}_{-0.05}$ \\
$v\sin(i) = 21.9^{+1.1}_{-1.1}$ km\,s$^{-1}$&
$\epsilon = -0.58^{+0.10}_{-0.08}$ dex\\
\hline
\end{tabular}\label{VegaRes}
\vspace{4mm}
\tabcolsep 2.6mm
\renewcommand{\arraystretch}{1}
\caption{Vega: Comparison of the results.}
\begin{tabular}{clcclcc}
\hline\hline
           & model     & \te  & \lg 
& \multicolumn{1}{c}{$\epsilon$} & $\xi_t$ & \vs \\
&& K&&& \kms & \kms \\
\hline
1) &  ATLAS\,6  & 9\,400 & 3.95 &   ~~~0.0 & 2.0\\
2) &  MARCS     & 9\,650 & 3.90 &   ~~~0.0 & 3.0\\
3) &  ATLAS\,6  & 9\,500 & 3.90 &   ~~~0.0 & 2.0\\
4) &  ATLAS\,6  & 9\,500 & 3.90 & $-$0.5 & 2.0\\
5) &  ATLAS\,12 & 9\,550 & 3.95 & $-$0.5 & 2.0\\
6) &  ATLAS\,9  & 9\,506 & 4.00 & $-$0.6 & 1.1 & 21.9\\
7) &  LLmodels  & 9\,540 & 3.92 & $-$0.58 & 2.4& 21.9\\
\hline
\end{tabular}\vspace{1mm}\\
1) Kurucz (1979), 2) Dreiling \& Bell (1980), 3) Lane \& Lester (1984),
4) Gigas (1986), 5) Castelli \& Kurucz (1994), 6) Hill et al. (2004), 7) our result
\label{VegaComp}
\end{table}

\begin{table}
\caption{Reduction in the errors of the parameters in \% by fixing one of them.}
\begin{tabular}{ccccc}
\hline
\hline
$\epsilon$ & $\xi_t$ & $\log(g)$ & $T_{\rm eff}$ & $v\sin(i)$\\
\hline
fixed &  86   &  \bf 67   &  \bf 64   &  98\\
100   & fixed & 100   & 100   & 100\\
\bf 56   &  81   & fixed &  \bf 39   &  77\\
\bf  56   &  78   &  \bf 50   & fixed &  95\\
100   &  97   & 100   &  98   & fixed\\
\hline
\end{tabular}\label{VegaErr}
\end{table}

\begin{figure}
\epsfig{figure=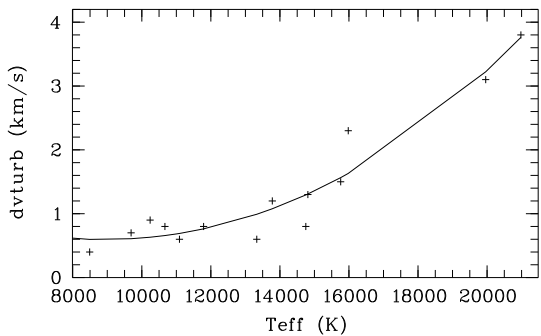, angle=-90, width=7cm, clip=}
\caption{Error in $\xi_t$ versus measured \te.}
\label{Xi_err}
\end{figure}

\begin{table*}
\caption{Fundamental parameters. $T_{\rm eff}^{\rm K}$ and $\log{g}^{\rm K}$ are taken from the KIC and 
given for comparison. Stars are sorted by their KIC number.}
\renewcommand{\arraystretch}{1.4}
\tabcolsep 1.97mm
\begin{tabular}{|r|rrc|ccc|cc|rc|lll|}
\hline
\multicolumn{1}{|c}{KIC} & 
\multicolumn{1}{|c}{$T_{\rm eff}^{\rm K}$} & 
\multicolumn{1}{c}{$T_{\rm eff}$} & 
$\frac{\d 1\sigma}{\d T_{\rm eff}}$ & $\log{g}^{\rm K}$ & $\log{g}$ & $1\sigma$ & 
$\xi_t$ & $1\sigma$ & $v\sin{i}$ & $\frac{\d 1\sigma}{\d v\sin{i}}$ & \multicolumn{3}{c|}{spectral type}\\
 &  \multicolumn{2}{c}{K} & \% & & & & \multicolumn{2}{c|}{km\,s$^{-1}$} & km\,s$^{-1}$ & \% & CDS & KIC & new\\   
\hline\hline
3\,240\,411~~       &	       & $20\,980_{-840}^{+880}$       & 4.1 &        & $4.01_{-0.11}^{+0.12}$ & 0.11 & $4.8_{-4.7}^{+2.9}$  	  & 3.8 & $42.6_{-4.9}^{+5.1}$  & 11  & --    & --      & B2\,V     \\
\hline
3\,756\,031~~       & 11\,177  & $15\,980_{-300}^{+310}$       & 1.9 & 4.24   & $3.75_{-0.06}^{+0.06}$ & 0.06 & $0.5_{-0.5}^{+2.3}$  	  & 2.3 & $30.8_{-3.1}^{+3.8}$  & 11  & --    & B8.5\,V & B5\,V-IV \\
\hline
5\,130\,305~~       &  9\,533  & $10\,670_{-200}^{+180}$       & 1.8 & 4.14   & $3.86_{-0.07}^{+0.07}$ & 0.07 & $1.4_{-1.0}^{+0.7}$  	  & 0.8 & $155_{-13}^{+13}$	& 8.4 & B9    & A0\,V   & B9\,V-IV\\
\hline
5\,217\,845$^{1)}$  &  8\,813  & $11\,790_{-260}^{+240}$       & 2.1 & 3.70   & $3.41_{-0.08}^{+0.10}$ & 0.09 & $2.1_{-0.9}^{+0.8}$  	  & 0.8 & $237_{-16}^{+16}$	& 6.8 & B8    & A3\,IV  & B8.5\,III   \\
\hline
5\,479\,821~~       & 10\,850  & $14\,810_{-290}^{+350}$       & 2.2 & 4.19   & $3.97_{-0.07}^{+0.09}$ & 0.08 & $0.1_{-0.1}^{+1.3}$  	  & 1.3 & $ 85_{-8}^{+8}$	& 9.4 & B8    & B9\,V   & B5.5\,V   \\
\hline
7\,599\,132~~       & 10\,251  & $11\,090_{-140}^{+100}$       & 1.1 & 3.62   & $4.08_{-0.06}^{+0.06}$ & 0.06 & $1.6_{-0.6}^{+0.5}$  	  & 0.6 & $ 63_{-4}^{+5}$	& 7.1 & B9    & B9.5\,IV & B8.5\,V     \\
\hline
8\,177\,087~~       &  9\,645  & $13\,330_{-170}^{+220}$       & 1.5 & 4.10   & $3.42_{-0.06}^{+0.06}$ & 0.06 & $1.3_{-0.7}^{+0.5}$  	  & 0.6 & $ 22.2_{-1.7}^{+1.5}$ & 7.2 & B9    & A0\,V  & B7\,III   \\
\hline
8\,389\,948~~       &  8\,712  & $10\,240_{-220}^{+340}$       & 2.7 & 3.61   & $3.86_{-0.10}^{+0.12}$ & 0.11 & $0.8_{-0.8}^{+0.9}$  	  & 0.9 & $ 142_{-11}^{+12}$	& 8.1 & B9\,V & A3\,IV & B9.5\,V-IV   \\
\hline
8\,451\,410$^{2)}$      &  8\,186  &  $8\,490_{-100}^{+100}$       & 1.2 & 3.81   & $3.51_{-0.05}^{+0.07}$ & 0.06 & $3.8_{-0.4}^{+0.4}$  	  & 0.4 & $39.8_{-1.4}^{+1.4}$  & 3.5 & B9\,V & A5\,IV & A3.5\,IV-III   \\
\hline
8\,459\,899$^{3)}$  &  9\,231  & $15\,760_{-210}^{+240}$       & 1.4 & 4.22   & $3.81_{-0.05}^{+0.05}$ & 0.05 & $1.4_{-1.4}^{+1.6}$  	  & 1.5 & $53_{-4}^{+4}$	& 7.5 & B8    & A1\,V  & B4.5\,IV \\
\hline
8\,583\,770~~       &  7\,659  & $ 9\,690_{-170}^{+230}$       & 2.1 & 3.47   & $3.39_{-0.05}^{+0.08}$ & 0.07 & $1.3_{-0.8}^{+0.6}$  	  & 0.7 & $ 102_{-7}^{+9}$	& 7.8 & B9    & A7\,IV-III& A0.5\,IV-III\\
\hline
8\,766\,405$^{3)}$  & 10\,828  & $12\,930_{-220}^{+210}$       & 1.7 & 3.67   & $3.16_{-0.08}^{+0.08}$ & 0.08 & $0.0^{+1.2}$	     	  & 1.2 & $240_{-12}^{+12}$	& 5.0 & B8    & B9\,IV   &  B7\,III \\
\hline
10\,960\,750~~      &	       & $19\,960_{-880}^{+880}$       & 4.4 &        & $3.91_{-0.11}^{+0.11}$ & 0.11 & $0.0^{+3.1}$	     	  & 3.1 & $253_{-15}^{+15}$	& 5.9 & B8    & --       & B2.5\,V   \\
\hline
11\,973\,705$^{4)}$ &  7\,404  & \multicolumn{1}{l}{(11\,150)} &     & (4.04) &\multicolumn{1}{l}{(3.96)} &   & \multicolumn{1}{l}{(3.9)} &     & $103_{-10}^{+10}$     & 9.7 & B9    & A9\,V-IV & B8.5\,V-IV \\
\hline
12\,207\,099$^{3)}$ & 10\,711  & \multicolumn{1}{l}{($<$11\,000)} &  & 4.07   &\multicolumn{1}{l}{$<$3.1} &   & \multicolumn{1}{l}{(2.5)} &     & $43_{-3}^{+5}$	& 9.3 & A0    & B9\,V    & B9\,III--II \\
\hline
12\,258\,330~~      & 13\,224  & $14\,700_{-200}^{+200}$       & 1.4 & 4.86   & $3.85_{-0.04}^{+0.04}$ & 0.04 & $0.0^{+0.8}$              & 0.8 & $130_{-8}^{+8}$	& 6.2 & B9\,V & B7\,V    & B5.5\,V-IV     \\
\hline
\end{tabular}\\
$^{1)}$binary, $^{2)}$suspected SB2, RV var., $^{3)}$suspected SB2, $^{4)}$SB2
\label{ResPar}
\end{table*}

\begin{table*}
\tabcolsep 3mm
\caption{Metallicity and elemental abundances relative to solar ones in dex and He abundance in fractions of the solar one. Stars are sorted by \te.}
\begin{tabular}{rrcccccccccc}
\hline\hline
\multicolumn{1}{c}{KIC} &
\multicolumn{1}{c}{\te} 
                               & $\epsilon$        & He     &	C	&  N	    &	O     &   Mg	  &	   Si &   S	  &   Ca    & Fe      \\
\hline
 8\,451\,410~~      &  8\,490  &$+0.10$$\pm$0.10   & --     & --	& --	    &$+$0.03   &$+$0.11   &$+$0.03    &$-$0.05    &$-$0.47  &$+$0.34   \\
 8\,583\,770~~      &  9\,690  &$+0.18$$\pm$0.09   &0.89    & --	& --	    &$+$0.18   &$+$0.31   &$+$0.13    &$-$0.20    &$+$0.23  &$+$0.24   \\
 8\,389\,948~~      & 10\,240  &$+0.10$$\pm$0.12   &1.24    & --	& --	    &$+$0.23   &$+$0.26   &$+$0.28    &$\pm$0.00  &$\pm$0.00&$+$0.14   \\
 5\,130\,305~~      & 10\,670  &$-0.07$$\pm$0.11   &1.13    & --	& --	    &$+$0.03   &$-$0.04   &$+$0.03    &$-$0.05    &$-$0.37  &$+$0.09   \\
12\,207\,099$^{1)}$ &$<$11\,000&($>$0.8)	   &1.40    & --	& --	    &$\pm$0.00 &$+$0.56   &$-$0.47    &$-$0.10    &$-$0.62  &$+$0.49   \\
 7\,599\,132~~      & 11\,090  &$+0.06$$\pm$0.10   &0.89    & --	& --	    &$+$0.23   &$+$0.16   &$+$0.13    &$\pm$0.00  &$-$0.07  &$+$0.09   \\
11\,973\,705$^{1)}$ & 11\,150  &~~(0.00$\pm$0.12)  &0.77    & --	& --	    &$+$0.33   &$+$0.21   &$-$0.57    &$-$0.40    &$+$0.43  &$+$0.04   \\
 5\,217\,845~~      & 11\,790  &$-0.06$$\pm$0.10   &1.08    &$+$0.15	&$\pm$0.00  &$\pm$0.00 &$+$0.31   &$+$0.03    &$+$0.05    &$+$0.38  &$-$0.11   \\
 8\,177\,087~~      & 13\,330  &$-0.11$$\pm$0.11   &\bf1.70 &$+$0.20	&$-$0.10    &$+$0.23   &$+$0.06   &$-$0.12    &$-$0.15    &$-$0.17  &$-$0.11   \\
 8\,766\,405~~      & 12\,930  &$-$\bf0.41$\pm$0.12&1.11    & --        &$+$\bf1.10 &$+$\bf1.11&$-$0.09   &$-$\bf0.37 &$-$\bf0.30 &    --   &$-$\bf0.56\\
12\,258\,330~~      & 14\,700  &$-$\bf0.30$\pm$0.16&\bf 2.10&$\pm$0.00  &$\pm$0.00  &$-$0.07   &$-$0.14   &$-$0.03    &$-$0.30    &    --   &$-$0.16   \\
 5\,479\,821~~      & 14\,810  &($-0.11$$\pm$0.15) &\bf 0.46&\bf $+$0.95&\bf $+$1.56&$\pm$0.00 &$-$\bf0.64&$+$0.13    &$-$\bf0.40 &    --   &$+$0.29   \\
 8\,459\,899~~      & 15\,760  &$-$\bf0.45$\pm$0.11&1.47    &$+$0.05	&$\pm$0.00  &$+$\bf0.38&$-$\bf0.54&\bf$-$0.47 &$-$\bf0.35 &    --   &$-$\bf0.46\\
 3\,756\,031~~      & 15\,980  &$-$\bf0.57$\pm$0.08&1.49    &$-$0.20	&$-$0.09    &$+$\bf0.38&$-$\bf0.44&\bf $-$0.67&\bf $-$0.60&    --   &$-$\bf0.36\\
10\,960\,750~~      & 19\,960  &$-0.04$$\pm$0.16   &1.00    &$\pm$0.00  &$\pm$0.00  &$\pm$0.00 &$\pm$0.00 &$\pm$0.00  &$\pm$0.00  &    --   &$\pm$0.00\\
 3\,240\,411~~      & 20\,980  &$-$\bf0.30$\pm$0.14&1.40    &$-$\bf0.30	&$+$0.06    &$-$0.12   &$-$0.19   &\bf $-$0.92&\bf $-$0.55&    --   &$-$\bf0.46\\
\hline
\multicolumn{1}{c}{KIC} &
\multicolumn{1}{c}{\te} 
                                &&   Na      &   Sc	 &   Ti  &   Cr  &   Mn      &   Y	 &  Ba       &&\\
\hline
 8\,451\,410~~     &  8\,490    &&\bf $-$0.58&\bf $-$2.01&$-$0.11&$+$0.40&$+$0.25    &\bf $+$0.93&\bf $+$1.57&&\\
 8\,583\,770~~     &  9\,690    &&  --       &$\pm$0.00  &$+$0.24&$+$0.30&\bf $+$0.50&\bf $+$1.13& --	     &&\\ 
 8\,389\,948~~     & 10\,240    &&  --       &$\pm$0.00  &$-$0.26&$-$0.20&  --       &  --	 & --	     &&\\
 5\,130\,305~~     & 10\,670    &&  --       &  --	 &$-$0.01&$-$0.05&  --       &  --	 & --	     &&\\
12\,207\,099$^{1)}$&$<$11\,000  &&  --       &  --	 &$-$0.16&$+$2.30&$\pm$0.00  &  --	 & --	     &&\\ 
 7\,599\,132~~     & 11\,090    &&  --       &  --	 &$+$0.09&$+$0.10&  --       &  --	 & --	     &&\\
\hline
\end{tabular}\\
$^{1)}$formal solution, suspected SB2 star
\label{ResAbund}
\end{table*}

Table\,\ref{VegaErr} gives the fraction of the computed error of each parameter in the case of
fixing one of the parameters to the error computed when all parameters are considered to be free. 
The values indicating the largest effects are set in bold face.
It shows that there is a 
strong correlation between \te, \lg, and metallicity. This can be easily understood 
because the line strengths are determined  by \te\ and $\epsilon$
and the shape of the Balmer line is strongly influenced by \lg\ and $\epsilon$. 
\vs\ and $\xi_t$, on the other hand, are much less influenced by the other parameters
and fixing them to their optimum values in the error calculation has no effect on the
derived errors of the other parameters. 
The result shows that we have to vary at least \te, \lg, and metallicity together to get reliable error
estimations for each single parameter.

\subsection{Results}
Table\,\ref{ResPar}  lists the fundamental parameters obtained for the 16 target stars.
\te\ and \lg\ taken from the KIC are given for comparison. The 1$\sigma$-errors of \te\ and \vs\ are given in 
per cent of the determined values, all other errors are absolute values.
The spectral types are compared between those given in the SIMBAD database at CDS, those derived from the 
\te\ and \lg\ listed in the KIC, and from our values of \te\ and \lg. For the derivation we used
an interpolation based on the tables by Schmidt-Kaler (\cite{Schmidt}) and by de Jager \& Nieuwenhuijzen 
(\cite{Jager}).

The accuracy in $T_{\rm eff}$ is about 2\% (mean value of 1.8\%) for most of the stars, 
only for the two hottest stars we obtain twice this value.
Looking for any correlations between the errors of measurement and the absolute values
of determined parameters, we found only one, namely for the micro-turbulent velocity in dependence
on \te\ (Fig.\,\ref{Xi_err}). It seems that our method can determine $\xi_t$ with an 
accuracy of about 0.6\,\kms for the cooler stars but that the error raises up to almost 4\,\kms\
for the hottest stars. It means that for stars hotter than about 15\,000\,K it is not possible to 
determine their micro-turbulent velocity and that for those stars the determination of the other 
parameters is practically independent of the value of $\xi_t$. There is also a slight
correlation between the relative error of \vs\ and \lg\ that we attribute to the fact 
that the accuracy of \vs\ is lowered by the inclusion of H$_{\beta}$ into the measurement and that
for lower \lg\ the Balmer lines show narrower profiles. Due to the inclusion of H$_{\beta}$, the mean
accuracy of our \vs\ determination is of only 8\%. The mean error in \lg\ is of 0.07. We refer to 
the previous section where it was shown that all the errors get larger (and more reliable)
by determining them from the complete grid in all parameters as we did here.

Table\,\ref{ResAbund} lists the resulting metallicities and individual abundances. 
The upper part gives the abundances of elements that were found in the spectra of most of the stars, 
the lower part those that have been additionally found for the cooler stars.
Larger deviations from the solar values are highlighted in bold face.
The solar values refer to the solar chemical composition given by Grevesse et al. (\cite{Grevesse}).
In the following, we give remarks on 
peculiarities found in some of the stars.  A general comparison of our results with the KIC data
will be given in Sect.\,6.2.

{\it KIC\,3\,240\,411 and 10\,960\,750:} Not any data about these two stars from previous investigations could be found.
According to our measurement, they are the hottest stars of our sample. KIC\,3\,240\,411 has low metallicity,
in agreement with a distinct depletion in Fe, whereas KIC\,10\,960\,750 has solar abundance.

{\it KIC\,3\,756\,031:} This star shows the lowest metallicity of our sample. All metals where we found
contributions in its spectrum are depleted, whereas He is enhanced.
The profiles of some of the metal lines are strongly asymmetric. This may indicate pulsation
(we will show that the star falls into the SPB instability region) or, because of the chemical
peculiarity of the star, rotational modulation due to spots
(as found, e.g., by Briquet et al. (\cite{Briquet}) for four B-type stars showing inhomogeneous
surface abundance distributions).

{\it KIC\,5\,217\,845:} The star has been identified from the Kepler satellite light curve as 
eclipsing binary of uncertain type with a period of 1.67838 days (\cite{2010arXiv1006.2815P}).
No RV or line profile variations could be detected from our two spectra, however.

{\it KIC\,5\,479\,821:} The star is He-weak by a factor of 0.46 compared to the solar value. It
shows a strong enhancement of C and N and a depletion of Mg and S. All final parameter values have been 
calculated from atmosphere models based on the derived individual abundances. The given metallicity,
derived from a previous step, has no physical meaning.

{\it KIC\,8\,177\,087:} This is the sharpest-lined star of our sample with \vs=22\,\kms. Whereas He is
strongly enhanced, all metal abundances are close to the solar values.

{\it KIC\,8\,451\,410:} This is a suspected SB2 star. We observe a RV shift between our two spectra 
and cannot fit the shifted and co-added spectrum as perfect as in the other cases. The observed strong
Ba and Y overabundance and the strong depletion in Sc may be related to the presence of a 
second component in the observed spectrum. The determined 
low temperature of about 8\,500\,K agrees within the errors of measurement with that given in the KIC.

{\it KIC\,8\,459\,899:} We suspect, from the slightly larger O-C residuals of the spectrum fit,
that this star may be a SB2 star as well but that the obtained parameters are still reliable.
It has low metallicity, the derived value agrees well with the depleted metal abundances, except for
O which is enhanced.

{\it KIC\,8\,583\,770:} This is a double star (WDS 19570+4441) with a 3-magnitude fainter companion
at a separation of 0.9\arcsec. According to our analysis, it is Si-strong, also Mg is enhanced.

{\it KIC\,8\,766\,405:} The star has low metallicity, the derived value agrees well with the depleted metal 
abundances, except for N and O which are enhanced.

{\it KIC\,11\,973\,705:} This is certainly a SB2 star, the typical spectrum of a cooler star of 8\,000 to 9\,000\,K
can be seen in the O-C residuals. All values given here result from a formal solution for 
the hotter component that shows the stronger lines. Due to the unknown flux ratio between the two components, 
the derived values, except for \vs, are not reliable.

{\it KIC\,12\,207\,099:} For this sharp-lined star we have got no satisfactory solution. Maybe that it is a 
Chromium star (we obtain a Cr abundance of 2.30 dex above the solar one) and 
vertical stratification of the elemental abundances has to be included into the calculations, 
or it is a SB2 star as well.

{\it KIC\,12\,258\,330:} The star is He-strong, it shows an enhancement in He abundance by a factor 
of more than two compared to the solar value. The metal abundances are close to the solar values, the derived
metallicity is too low (see the remark for KIC\,5\,479\,821).

\section{NLTE based analysis}

Our aim in this attempt was not to investigate the NLTE effects in detail but to 
check for the order of the deviations in the results of the LTE calculations that may arise from neglecting these effects.
We used the GIRFIT program (Fr\'emat et al. \cite{Fremat06}) to determine the fundamental parameters \te, \lg,
and \vs. This program adjusts synthetic spectra interpolated on a grid of stellar fluxes to the observed spectra
using the least squares method. 
The temperature structure of the atmospheres was computed as in Castelli
\& Kurucz (\cite{Castelli03}) using the ATLAS9 computer code (Kurucz\,\cite{KuruczA}). Non-LTE level populations were then computed
for each of the atoms listed in Table\,\ref{table:atoms} using the TLUSTY program (Hubeny \& Lanz \cite{Hubeny}) and
keeping the temperature and the density distributions obtained with ATLAS9 fixed. For the spectral region
considered, we used the specific intensity grids computed by Fr\'emat (private communication) for \te\ and \lg\
ranging from 15\,000 K to 27\,000 K and from 3.0 to 4.5,
respectively. For \te$<$15000\,K we used LTE calculation. We assumed solar metallicity and a
micro-turbulence of 2\,\kms\ for all the grids.

\begin{table}
\caption{Atomic models used for the treatment of NLTE.}
\begin{tabular}{lll}
\hline\hline
atom & ion & levels\\
\hline
Hydrogen  & \ion{H}{I}     &  8 levels + 1 superlevel\\
          & \ion{H}{II}    &  1 level\\
Helium    & \ion{He}{I}    & 24 levels\\
          & \ion{He}{II}   & 20 levels\\
          & \ion{He}{III}  &  1 level\\
Carbon    & \ion{C}{II}    & 53 levels, all individual levels\\
          & \ion{C}{III}   & 12 levels\\
          & \ion{C}{IV}    &  9 levels + 4 superlevels\\
          & \ion{C}{V}	   &  1 level\\
Nitrogen  & \ion{N}{I}	   & 13 levels\\
          & \ion{N}{II}    & 35 levels  + 14 superlevels\\
          & \ion{N}{III}   & 11 levels\\
          & \ion{N}{IV}    &  1 level\\
Oxygen    & \ion{O}{I}	   & 14 levels  + 8  superlevels\\
          & \ion{O}{II}    & 36 levels  + 12 superlevels\\
          & \ion{O}{III}   &  9 levels\\
          & \ion{O}{IV}    &  1 level\\
Magnesium & \ion{Mg}{II}   & 21 levels  + 4 superlevels\\
          & \ion{Mg}{III}  &  1 level\\
\hline
\end{tabular}
\label{table:atoms}
\end{table}

\begin{table}
\caption{Fundamental parameters derived from the NLTE model. LTE based values 
taken from Table\,\ref{ResPar} are given in parentheses.}
\begin{tabular}{llcc}
\hline\hline
Kepler-ID       & \multicolumn{1}{c}{\te\ (K)}  & \lg & \vs\ (\kms) \\
\hline
KIC\,3\,240\,411      & ~20\,200$\pm$1000 &  4.0$\pm$0.2    & ~~~~45$\pm$15  \\  
                      & (20\,980$\pm$860) & (4.0$\pm$0.1)   & ~~(43$\pm$5)   \vspace{1mm}\\
KIC\,10\,960\,750     & ~18\,100$\pm$1000 &  3.7$\pm$0.2    &  ~~250$\pm$20  \\ 
                      & (19\,960$\pm$880) & (3.9$\pm$0.1)   & ~~(253$\pm$15) \vspace{1mm}\\
KIC\,3\,756\,031      & ~16\,100$\pm$800  &  3.8$\pm$0.2    & ~~~~40$\pm$15  \\
                      & (15\,980$\pm$300) & (3.75$\pm$0.06) & ~~(31$\pm$4)   \vspace{1mm}\\ 
KIC\,12\,258\,330     & ~15\,400$\pm$800  &  4.3$\pm$0.2    &  ~~125$\pm$15  \\
KIC\,12\,258\,330$^*$ & ~14\,700$\pm$800  &  4.1$\pm$0.2    &  ~~125$\pm$15  \\
                      & (14\,700$\pm$200) & (3.85$\pm$0.04) & (130$\pm$8)    \\
\hline
\end{tabular}\\
$^*$ based on H$_{\beta}$ only 
\label{table:results}
\end{table}

\begin{figure}
\epsfig{figure=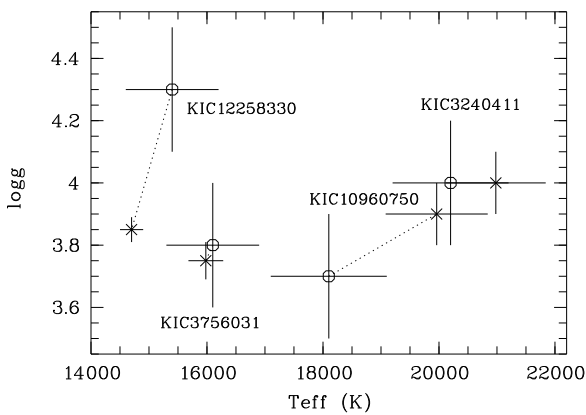, angle=-90, width=7.6cm, clip=}
\caption{Comparison of the parameter values derived from LTE (crosses) and from NLTE (circles) treatment.}
\label{CompTlgg}
\end{figure}

Our investigation of NLTE effects in the four hottest stars of our sample (not regarding the suspected SB2 candidates)
is based on the 4720-5050\,\AA\ spectral region. 
For the hotter stars, we used the two helium
lines at 4921 and  5016 \AA\ that are present in this region to measure \vs. Then we determined the other
parameters, \te\ and \lg, by fitting the spectrum in the whole domain between 4720 and 5050\,\AA.
For the cooler stars of the sample, the determination of \vs\ is based on the metallic and the He I
lines. The derived \vs\ values of these stars are consistent with those obtained for the $H_{\beta}$ lines.
The coolest star, KIC\,12\,258\,330, was analysed using LTE calculations to check for the influence of the usage of 
different wavelength ranges and the differences in the applied programs.

Table~\ref{table:results} lists the results.
The given errors are computed according to the error determination method introduced by
Martayan et al. (\cite{Martayan}) that is based on the computation of theoretical spectra where a Poisson distributed noise
has been added. 
For KIC\,12\,258\,330, we did not find a satisfying solution that fits both the Balmer and the He lines and give
two solutions here. 
For comparison, we also list the parameter values obtained in Sect.\,3 and visualize the 
differences in the results in Fig.\,\ref{CompTlgg}.

For two of the stars, among them the hottest star KIC\,3\,240\,411, the values of both \te\ and \lg\
agree within the errors of measurement. For the other two stars, surprisingly for KIC\,12\,258\,330 where we assumed
LTE, we observe large differences between the values obtained from the two approaches. 
These results will be discussed in detail in Sect.\,6.1.

\section{Stellar temperatures from spectral energy distributions}
\subsection{The method}

Stellar effective temperatures can be determined from the spectral energy distributions (SEDs).
For our target stars, these were constructed from photometry taken from the literature.
2MASS (\cite{2006AJ....131.1163S}),
Tycho B and V magnitudes (\cite{1997A&A...323L..57H}),
USNO-B1\,R magnitudes (\cite{2003AJ....125..984M}),
and TASS\,I magnitudes (\cite{2006PASP..118.1666D}),
supplemented with CMC14 $r'$ magnitudes (\cite{2002A&A...395..347E})
and TD-1 ultraviolet flux measurements (\cite{1979BICDS..17...78C}) where available.

The SED can be significantly affected by interstellar
reddening. We have determined the reddening from interstellar Na\,D lines present in our
spectra. For resolved multi-component interstellar Na\,D lines, the equivalent widths of the individual
components were measured using multi-Gaussian fits. The total \ebv\ in these cases is the sum of the
reddening per component, since interstellar reddening is additive
(\cite{1997A&A...318..269M}).
\ebv\ was determined using the relation given by these authors.  
Several of stars have $UBV$ photometry which allows
us to determine \ebv\ using the Q-method (\cite{1973IAUS...54..231H}).
For these stars, there is good agreement with the extinction obtained from the
Na\,D lines. The SEDs were de-reddened using
the analytical extinction fits of \cite{1979MNRAS.187P..73S} for the ultraviolet
and \cite{1983MNRAS.203..301H} for the optical and infrared.

The stellar $T_{\rm eff}$ values were determined by fitting solar-composition
(Kurucz \cite{KuruczA}) model fluxes to the de-reddened SEDs.
The model fluxes were convolved with photometric filter response
functions. A weighted Levenberg-Marquardt, non-linear least-squares fitting procedure was
used to find the solution that minimizes the difference between the observed and model
fluxes. Since \lg\ is poorly constrained by our SEDs, we
fixed \lg=4.0 for all the fits. 

\begin{table}
\tabcolsep 1.7mm
\caption{\ebv\ determined from the Na\,D lines and from the Q-method, and \te\ 
obtained from SED-fitting.}
\begin{tabular}{rlcll} \hline\hline
\multicolumn{1}{c}{KIC}      & 
\multicolumn{2}{c}{\ebv} &
\multicolumn{1}{c}{\te}      & SED Notes       \\
&         & Na\,D & QM\\
\hline\hline
 3\,240\,411 &  0.07$\pm$0.01	  &	 & 22\,280$\pm$1\,320 & CMC14 $r'$    \\
 3\,756\,031 &  0.12$\pm$0.01$^*$ &	 & 18\,470$\pm$~~~970 & CMC14 $r'$    \\
 5\,130\,305 &  0.09$\pm$0.01	  &	 & 11\,590$\pm$~~~470 & CMC14 $r'$    \\
 5\,217\,845 &  0.25$\pm$0.03	  &	 & 18\,780$\pm$2\,250 & CMC14 $r'$    \\
 5\,479\,821 &  0.24$\pm$0.03	  &	 & 25\,280$\pm$3\,390 & 	      \\
 7\,599\,132 &  0.02$\pm$0.01	  &	 & 10\,300$\pm$~~~130 & CMC14 $r'$, TD1 \\
 8\,177\,087 &  0.12$\pm$0.01	  & 0.09 & 13\,120$\pm$~~~200 & TD1	     \\
 8\,389\,948 &  0.20$\pm$0.02	  & 0.19 & 12\,270$\pm$~~~550 & CMC14 $r'$, TD1 \\
 8\,451\,410 &  0.04$\pm$0.01	  &	 &~~8\,560$\pm$~~~120 & CMC14 $r'$   \\
 8\,459\,899 &  0.13$\pm$0.01$^*$ & 0.16 & 14\,780$\pm$~~~310 & TD1	     \\
 8\,583\,770 &  0.38$\pm$0.06	  &	 & 16\,290$\pm$3\,440 & CMC14 $r'$    \\
 8\,766\,405 &  0.10$\pm$0.01$^*$ & 0.12 & 15\,460$\pm$~~~750 & 	     \\
10\,960\,750 &  0.06$\pm$0.01	  & 0.06 & 20\,530$\pm$~~~980 & CMC14 $r'$, TD1 \\
11\,973\,705 &  0.02$\pm$0.01$^*$ &	 &~~7\,920$\pm$~~~100 & TD1	      \\
12\,207\,099 &  0.03$\pm$0.01	  &	 & 12\,160$\pm$~~~520 & 	      \\
12\,258\,330 &  0.04$\pm$0.01	  & 0.08 & 15\,820$\pm$~~~370 & TD1	      \\
\hline
\end{tabular}
\label{Teff-SED}
$^*$ indicates multi-component interstellar Na\,D lines
\end{table}

\begin{figure}
\epsfig{figure=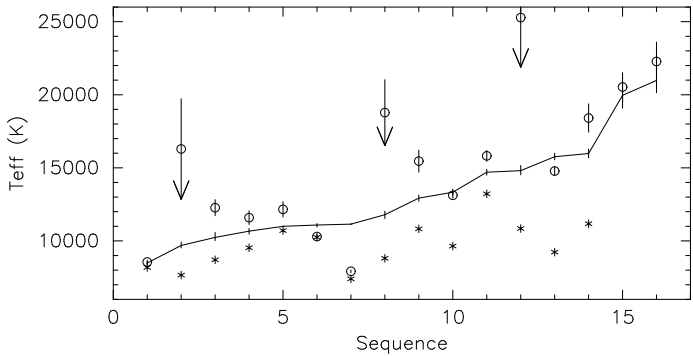, angle=-90, width=9cm, clip=}
\caption{Comparison of \te\ given in the KIC (asterisks) with the
spectroscopic (connected by lines) and photometric (open circles) values. 
The stars are sorted by their spectroscopically determined temperature.}
\label{Comp3}
\end{figure}

\subsection{Results}
The results are given in Table~\ref{Teff-SED}. The
uncertainty in $T_{\rm eff}$ includes the formal least-squares error and that from the
uncertainty in \ebv\ added in quadrature. The differences between the photometric,
spectroscopic, and the KIC values are shown in Fig.\,\ref{Comp3}. 

KIC\,10\,960\,750 has $uvby\beta$ photometry. Using the {\sc uvbybeta} and {\sc tefflogg}
codes of \cite{1985CommULO.....78M}, we obtain \ebv=0.05, \te=19320$\pm$800~K, and \lg=3.60$\pm$0.07, 
which is in good agreement with what we determined from spectroscopy. For three more stars, the 
\te\ derived from SED fitting agree with those from the spectroscopic analysis within
the errors of measurement. For 12 of the 16 targets, the photometric \te\ 
is close to the spectroscopic \te\ confirming the spectroscopically obtained values. In only one case,
for KIC\,11\,973\,705, the photometric \te\ is in favour of the KIC value. This is the star that we identified
as a SB2 star where we see the lines of a secondary component in its spectrum. 
KIC\,5\,217\,845, 5\,479\,821 and 8\,583\,770 suggest that the interstellar lines
comprise of un-resolved multiple components leading to an overestimation of the interstellar reddening
and to much too high \te. One of them, KIC\,8\,583\,770, is a double star (WDS 19570+4441) with a 3-magnitude fainter companion
at a separation of 0.9\arcsec.

\section{Discussion}

\subsection{The influence of NLTE effects}

Four stars have been analysed by two different methods using
a) slightly different input physics, b) LTE or NLTE treatment for three of the stars, and
c) different wavelength ranges, short in NLTE, wider in LTE.
For two of the stars, KIC\,3\,240\,411 and KIC\,3\,766\,081, the results agree
well within the errors of measurement. For the other two stars, KIC\,10\,960\,750 and
KIC\,12\,258\,330, we find significant deviations. 
The fact that we find a good agreement for the hottest star and a distinct deviation for the coolest
one that was analysed using LTE in both attempts, is surprising and raises questions about the origin
of the observed deviations, i.e. about the order of the influence of the differences in the two methods
that we labeled a) to c) in before. 

\begin{figure}
\epsfig{figure=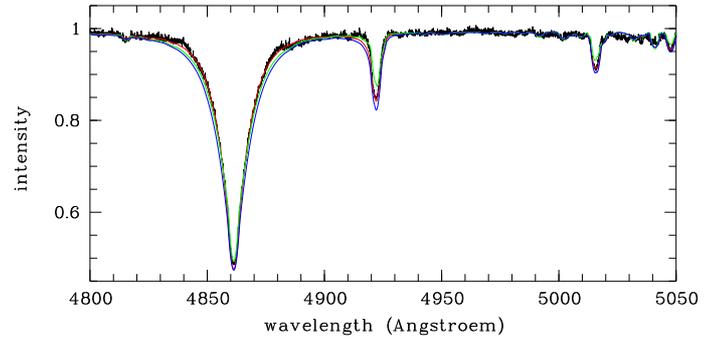, angle=-90, width=9cm, clip=}
\caption{Fit of the spectrum of KIC\,12\,258\,330 obtained from models 1 (red), 2 (green), and 3 (blue).}
\label{CompHe}
\end{figure}

For KIC\,12\,258\,330, we can clearly show that the difference in the results comes from the facts that 
it is a helium-strong star and that our second approach assumed solar metallicity and He abundance. 
Fig.\,\ref{CompHe} shows the quality of different fits, all calculated in LTE.
Model 1 is the original LTE solution 
with the He abundance enhanced by a factor of 2.1 against the solar one and \te=14\,700\,K, \lg=3.85.
Models 2 and 3 are based on \te=15\,400 K, \lg=4.30 with solar He abundance in model 2 and enhanced
He abundance in model 3.
It can be seen that varying the
He abundance effects not only the strength of the He lines but also the shape of the wings of \hb\ and 
so the derived \lg. Compared to solution 1), the reduced $\chi^2$ of solution 2) is 1.5 times higher
and that of solution 3) 3.2 times higher.
We assume that the parameter values derived in Sect.\,3 are the correct ones and that KIC\,12\,258\,330
is helium-strong.

A closer investigation of the application to KIC\,10\,960\,750 shows that the differences in \te\ and \lg\
mainly come from the usage of different wavelength regions. If we shorten the region for the LTE calculations 
to that used in NLTE we end up with  \te=(18940$\pm$840)\,K and \lg=3.78$\pm$0.08 which comes much closer to the 
results obtained from the NLTE calculations. Here, we have to solve the question if the difference in the 
results comes from the fact that one more stronger He line at 5876\,\AA\ was included in the wider spectral range used in
the LTE approximation which falsifies the LTE results due to additional NLTE effects, or if the difference simply
comes from the fact that a wider spectral range gives more accurate results which favours the LTE results.

As already mentioned in the introduction, Auer \& Mihalas (\cite{Auer}) state that the deviations in
the equivalent widths of the He lines due to the effects of departure from NLTE increases for B-type stars
with wavelength and can reach 30\%  for the 5\,876\,\AA\ line and more for redder lines. According to their
calculations, NLTE effects are negligible only for the He lines in the blue (up to \ion{He}{I} 4471\,\AA) but
produce deeper line cores for the He lines at longer wavelengths whereas the line wings remain essentially
unaffected. Hubeny \& Lanz (\cite{Hube2}) state as well that the core of strong lines and lines from minor ions 
will be most affected by departures from LTE which implies that the abundance of some species might be
overestimated from LTE predictions. Also the surface gravities derived from the Balmer line
wings tend to be overestimated.

Mitskevich \& Tsymbal (\cite{Mitsk}), on the other hand, computed model atmospheres of B-stars in LTE and NLTE and found no
remarkable differences. In particular the temperature inversion in the upper layers of the atmospheres, intrinsic to the
models of Auer \& Mihalas, is absent in their NLTE models and the departure coefficients for the first levels of H and He
in the upper layers are lower by three orders of magnitude. The authors believe that the reason for the discrepancy
in the results is the absence of agreement between radiation field and population levels in the program applied 
by Auer \& Mihalas. And there is a second point. The TLUSTY program (Hubeny \& Lanz \cite{Hubeny}) provides NLTE fully 
line-blanketed model atmospheres, whereas Auer \& Mihalas did not take the line blanketing into account.

In a more recent article, Nieva \& Przybilla (\cite{Nieva}) investigate the NLTE effects in
OB stars. Using ATLAS\,9 for computing the stellar atmospheres in LTE and the DETAIL and SURFACE programs to
include the NLTE level populations and to calculate the synthetic spectra, respectively, 
they compute the Balmer and \ion{He}{I,\,II} lines
over a wide spectral range and compare the results with those of pure LTE calculations. In the result, they obtained narrower
profiles of the Balmer lines in LTE compared to NLT for stars hotter than 30\,000\,K, leading to an overestimation of their
surface gravities. The calculations done for one cooler star of 20\,000\,K which is in the range of the hottest stars of our sample,
but for \lg\ of 3.0, did not show such effect but differences in the line cores, increasing from H$_{\delta}$ to \ha. 
The same they observed for the cooler star for the \ion{He}{I} lines. Many of them experience significant NLTE-strengthening,
in particular in the red, but without following a strong rule. So the strengthening of the \ion{He}{I}
5876\,\AA\ line is less then that of the 4922\,\AA\ and 6678\,\AA\ lines.

From the LTE calculations, we obtain solar metal and He abundance for KIC\,10\,960\,750.
Unfortunately, our actually available grid of NLTE synthetic spectra does not comprise the 
wavelength region of the \ion{He}{I}
5876\,\AA\ line and so we cannot reproduce the LTE calculations one by one. 
Since the quality of the fit of the observed spectrum obtained from the LTE and NLTE calculations is the same
we can not directly decide if the difference in the parameters comes from NLTE effects of the He 5876\,\AA\ line  
as discussed above. But since we do not observe any deviation in the results for the hottest star of our sample,
we believe that NLTE effects are only second order effects and cannot give rise to the deviations observed for 
KIC\,10\,960\,750.

\subsection{Comparison with the KIC data}
\begin{figure}
\epsfig{figure=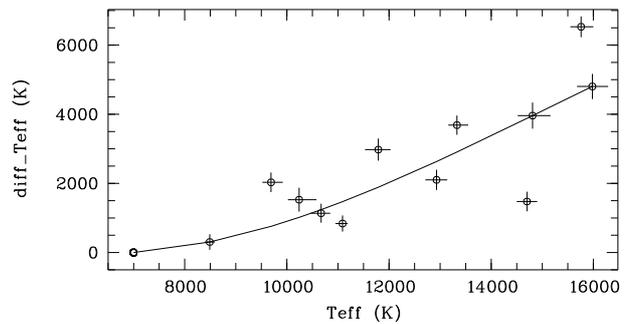, angle=-90, width=8cm, clip=}
\caption{Difference in \te\ between our values and those given in the KIC versus
measured \te.}
\label{Teff_KIC}
\end{figure}

Comparing our $T_{\rm eff}$ and $\log{g}$ with the values given for 12 of the stars in the KIC (for two of the 16 stars
of our sample there are no entries, and we excluded KIC\,11\,973\,705 
and 12\,207\,099), we see systematic differences. We obtain higher temperatures 
in general, the difference increases with increasing temperature of the stars. Already Molenda-\.Zakowicz et al. (\cite{Molenda}) found that
the $T_{\rm eff}$ given in the KIC is too low for stars hotter than about 7\,000\,K, by up to 4\,000\,K for the hottest stars.
The trend in the temperature difference derived from our values is shown in Fig.\,\ref{Teff_KIC} (the error in the difference
also includes the 200\,K error typical for the KIC data). We have drawn the curve 
of a second order polynomial resulting from a fit that observes the boundary condition that the difference should be zero for \te=7\,000\,K.
Our analysis seems to confirm the finding by Molenda-\.Zakowicz et al.

\begin{figure}
\epsfig{figure=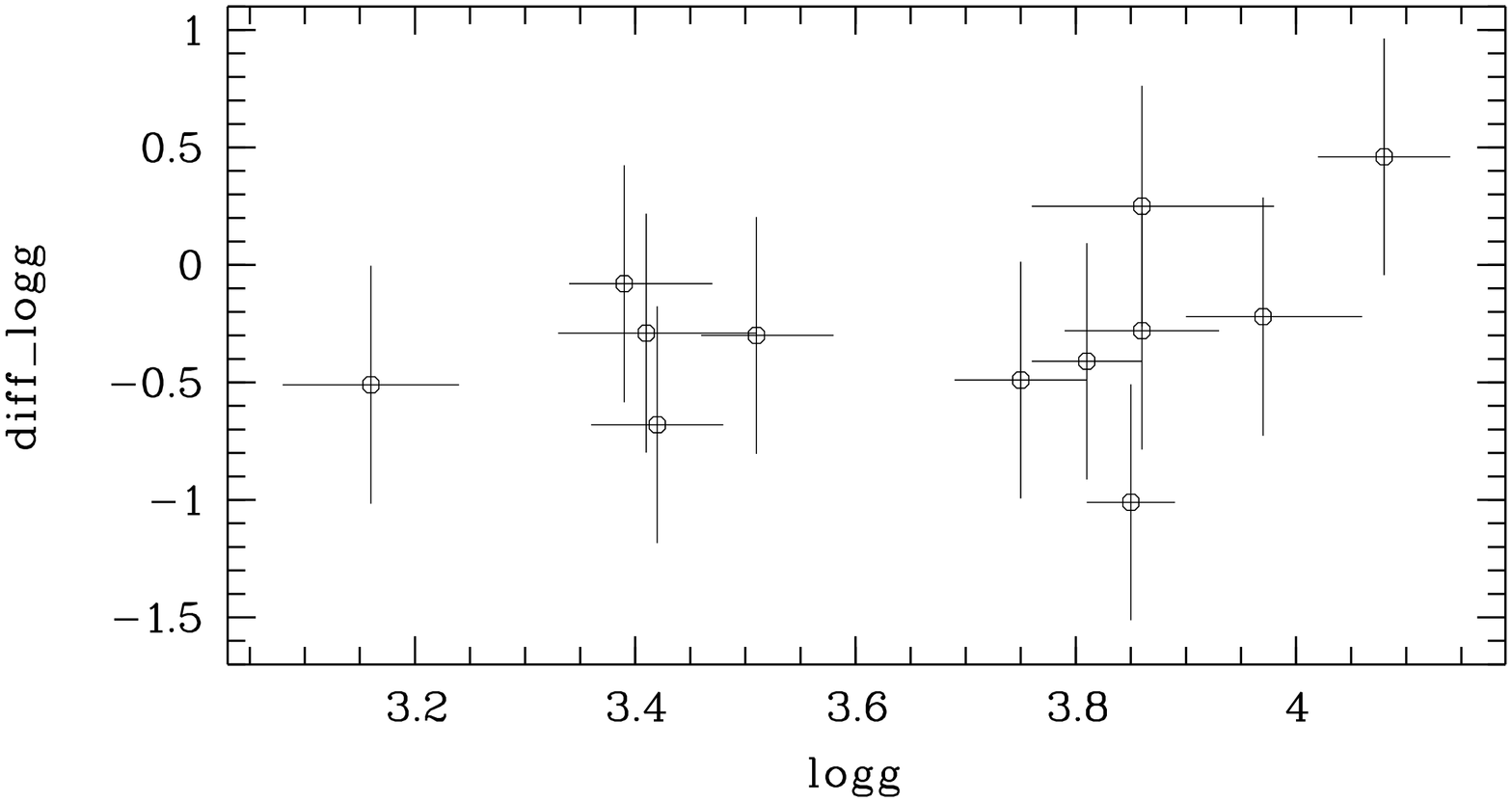, angle=-90, width=8cm, clip=}
\caption{Difference in \lg\ between our values and those given in the KIC versus
measured \lg.}
\label{logg_KIC}
\end{figure}

\begin{figure}
\epsfig{figure=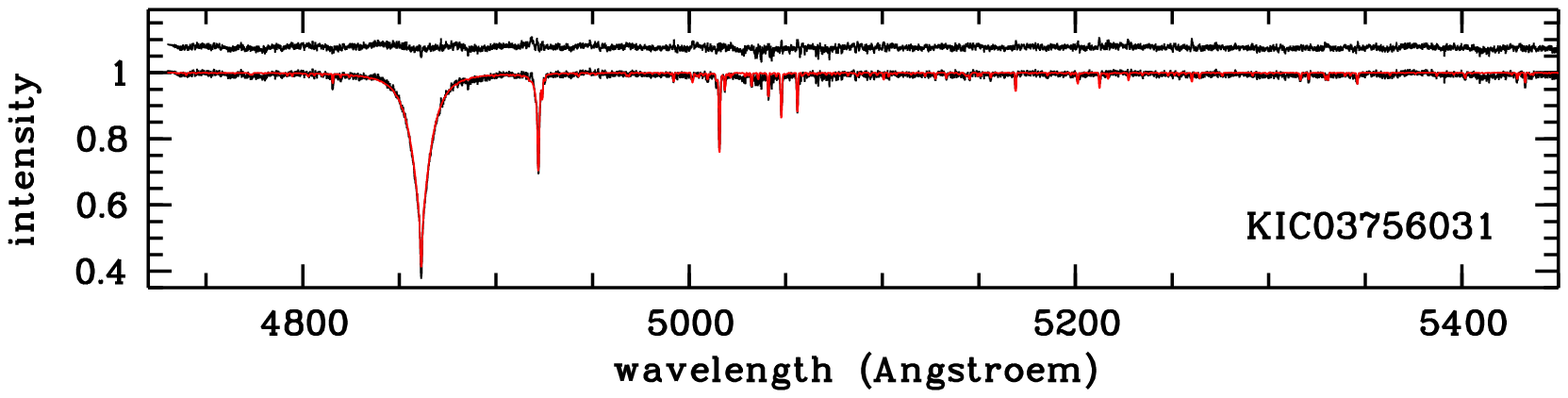, width=9cm, clip=}\vspace{-9mm}
\epsfig{figure=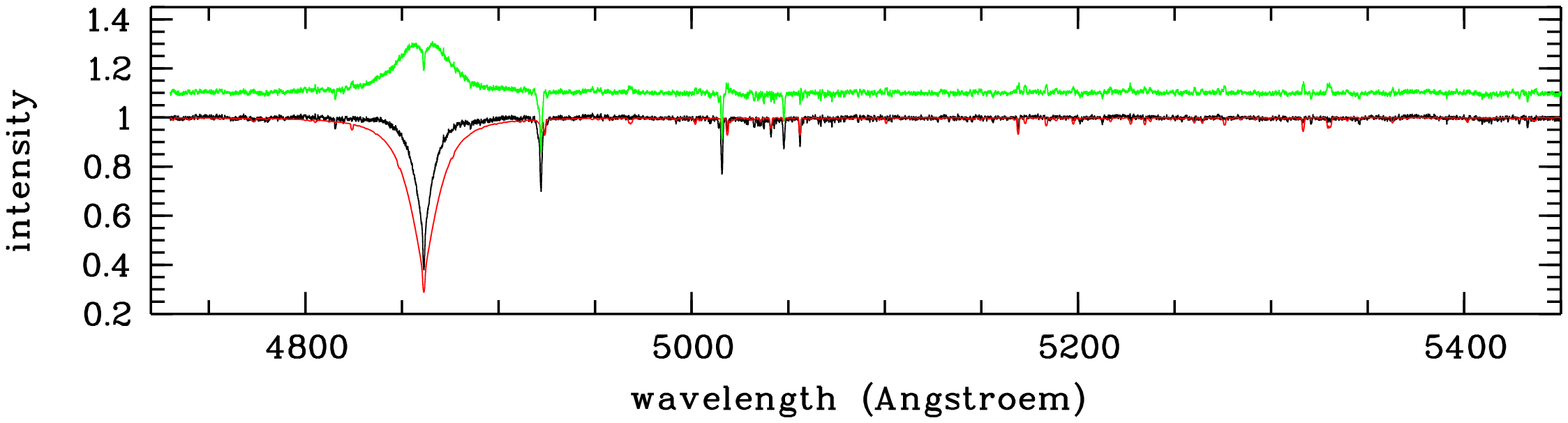, angle=-90, width=9cm, clip=}
\caption{Fit (red) of the spectrum of K03758031 (black), assuming \te=16\,000\,K (top)
and \te=11\,200\,K (bottom). The O$-$C spectra are shifted by +1.1.}
\label{K01}
\end{figure}

\begin{figure}
\hspace{-1mm}
\epsfig{figure=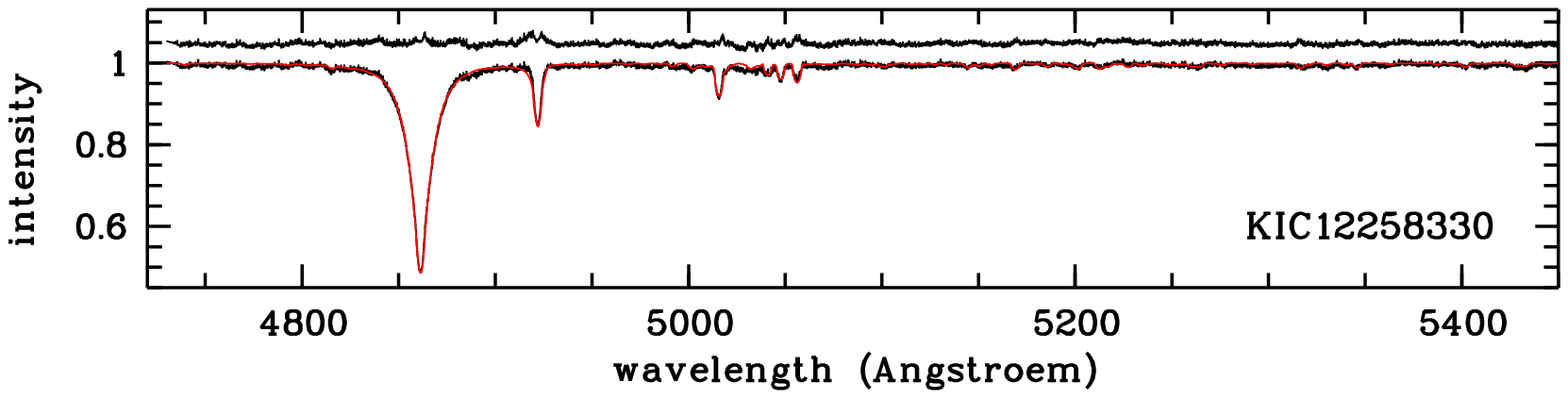, width=9cm, clip=}\vspace{-9mm}
\epsfig{figure=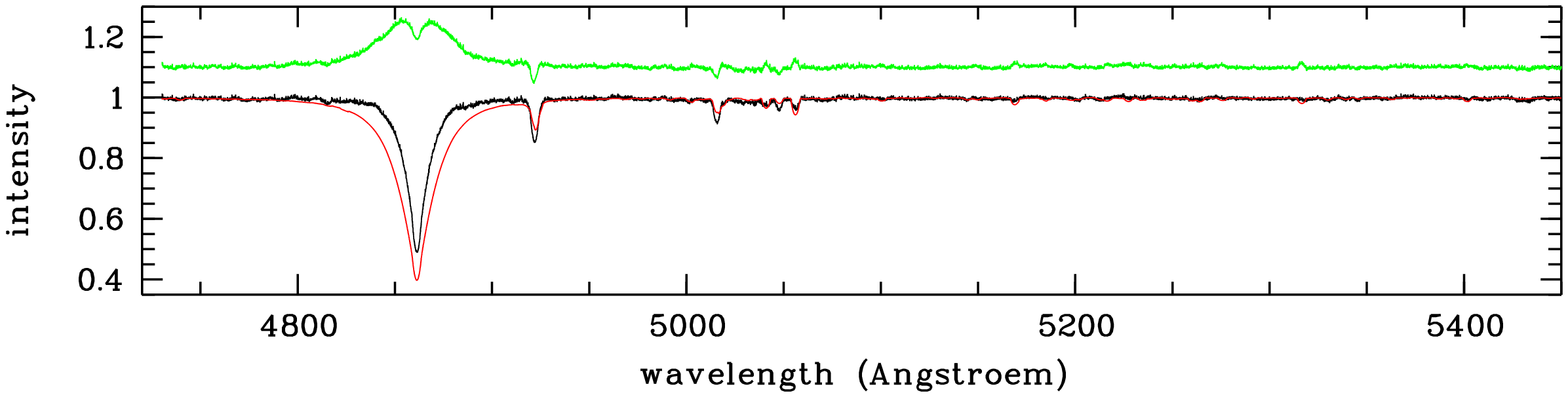, angle=-90, width=9cm, clip=}
\caption{ As Fig.\,\ref{K01} but for KIC\,12\,258\,330 and \te=14\,750\,K, \lg=3.85 (top) and
\te=13\,224\,K, \lg=4.86 (bottom).}
\label{K02}
\end{figure}

Comparing our values derived for \lg\ with those given in the KIC, we find that they are systematically lower
(Fig.\,\ref{logg_KIC}).
Counting for the large errors that mainly result from the $\pm$0.5\,dex error of the KIC data,
we cannot say that this difference is significant, however. The same holds true for the Fe/H
ratios given in the KIC, the corresponding error of $\pm$0.5\,dex prevents us from any comparison
with our derived values.

There are two extreme cases: KIC\,3\,756\,031, where 
our derived temperature is 16\,000\,K whereas the KIC gives about 11\,200\,K, and KIC\,12\,258\,330, where we obtain a
$\log{g}$ of 3.85 
which is by 1 dex lower than given in the KIC. A closer investigation shows, however, that the KIC values are very 
unlikely. 

Fig.\,\ref{K01} shows, in its upper panel, the fit of a part of the spectrum of KIC\,3\,756\,031 assuming our 
value of $T_{\rm eff}$. The lower panel gives the same for $T_{\rm eff}$ taken from the KIC. Neither H$_{\beta}$ nor the He lines are fitted well. 
The best fit of H$_{\beta}$ at this temperature is obtained assuming a $\log{g}$ below 3.0, 
but in this case the He lines can also not be reproduced. The use of LTE in our program cannot balance a temperature difference of 5\,000\,K and 
the 11200 K given in the KIC cannot be true for this star.
 
Fig.\,\ref{K02} shows, in its upper panel, the fit of a part of the spectrum of 
KIC\,12\,2583\,330 assuming our values of \te\ and \lg. The lower panel gives the same for \te\ and \lg\ 
taken from the KIC. The resulting 
deviation in the shape of H$_{\beta}$ cannot be explained in terms of a wrong continuum normalization in 
the H$_{\beta}$ range. The $\log{g}$ given for this star in the KIC is by about 1 dex too large.

\subsection{\te\ from spectral energy distributions}
Our method of deriving \ebv\ from the EWs of the Na\,D lines may overestimate \te\ in the cases where we observe 
non-resolved interstellar contributions to the Na\,D line profiles. This is one reason, besides the poor photometric
data for some of the stars, why this method is of lower accuracy compared to the spectroscopic analysis.
The facts that the \ebv\ derived from the Na\,D lines and from the Q-method for the stars where $UBV$ or, in one case, 
$uvby\beta$ photometry was available are in a good agreement and that all derived \te\ are in favour of
our spectroscopic values confirm that the \te\ given in the KIC must be too low, however. 

Thus, the results from the SED-fitting based on the photometric data reveal the reason why the \te\ given in the KIC
deviate from our findings. For most of the hotter stars, the interstellar 
reddening was not properly taken into account leading to an underestimation of the stellar temperatures. It also explains
why the difference in the derived temperatures between the KIC and our spectroscopic analysis rises with increasing
temperature of the stars. The hotter the stars, the more luminous and the farther they are and the more the 
ignored reddening plays a role.

\subsection{Positions of the stars compared to SPB and $\beta$\,Cep instability strips} 

\begin{figure}
\epsfig{figure=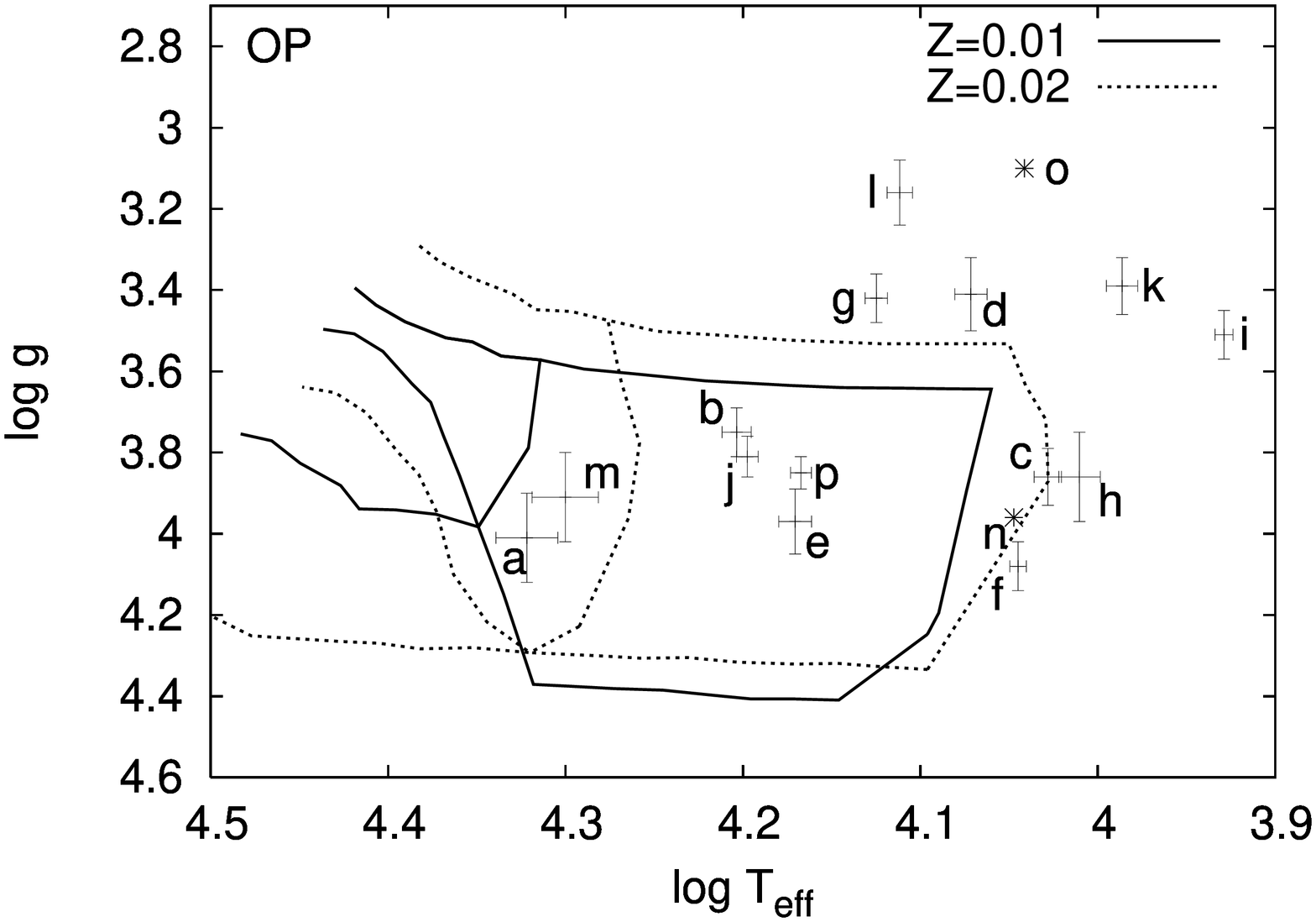, angle=-90, width=8cm, clip=}\vspace{1mm}
\epsfig{figure=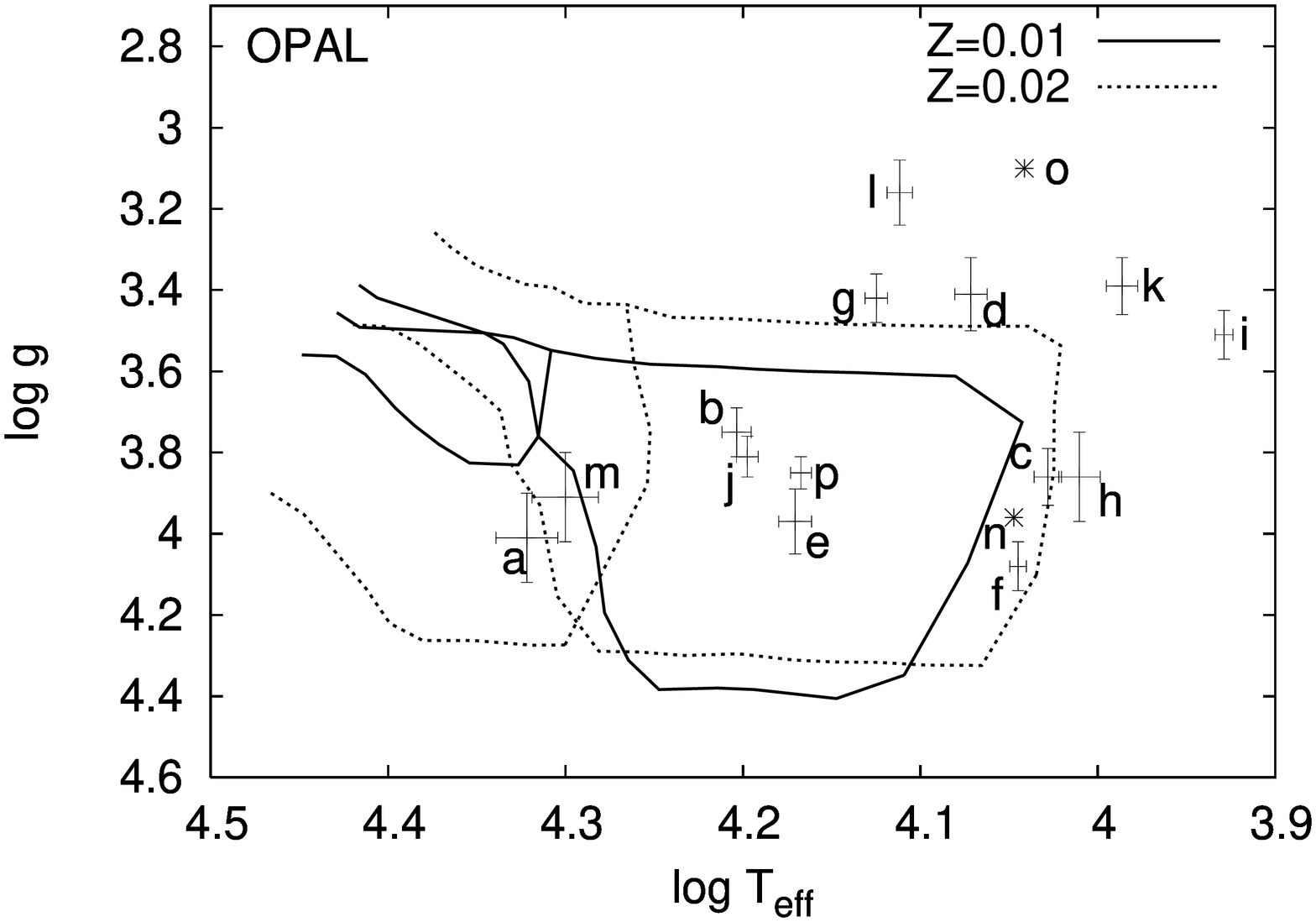, angle=-90, width=8cm, clip=}
\caption{Positions of the stars and the SPB and $\beta$\,Cep instability regions in the 
\te--\lg\ diagram.}
\label{logTeff_logg}
\end{figure}

\begin{table}
\caption{Positions of the stars with respect to the instability regions.}
\begin{tabular}{crlc}
\hline\hline
position & \multicolumn{1}{c}{KIC} & spectral type  & $N$\\
\hline
                                           &    3\,240\,411 & B2~~~\,V     & a\\
\raisebox{1.2ex}[-1.2ex]{$\beta$\,Cep/SPB} &     10\,960\,750 & B2.5\,V      & m\\
\hline    
                                           &    3\,756\,031 & B5~~~\,IV-V  & b\\	     
	                                   &    5\,479\,821 & B5.5\,V	   & e\\	 
\raisebox{1.2ex}[-1.2ex]{SPB}              &    8\,459\,899 & B4.5\,IV     & j\\
	                                   &     12\,258\,330 & B5.5\,IV-V   & p\\
\hline
                                           &    5\,130\,305 & B9~~~\,IV-V  & c\\
                                           &    5\,217\,845 & B8.5\,III    & d\\	  
	                                   &    7\,599\,132 & B8.5\,V	   & f\\
\raisebox{1.2ex}[-1.2ex]{possibly SPB}     &    8\,177\,087 & B7~~~\,III   & g\\
                                           &    8\,389\,948 & B9.5\,IV-V   & h\\
\hline
                                           &    8\,451\,410 & A3.5\,IV-III & i\\ 
\raisebox{1.2ex}[-1.2ex]{too cool}         &    8\,583\,770 & A0.5\,IV-III & k\\
\hline 
too evolved                                &    8\,766\,405 & B7~~~\,III   & l\\
\hline                                     
uncertain                                  & 11\,973\,705 & B8.5\,VI-V   & n\\
(SB2 stars)                                & 12\,207\,099 & B9~~~\,II-III    &  o\\ 
\hline	     
\label{pulsators}
\end{tabular}
\label{positions}
\end{table}

In the result of our analysis, we can directly place the stars into a \te--\lg\ 
diagram to compare their positions with the known instability domains of main-sequence B-type 
pulsators. Fig.\,\ref{logTeff_logg} shows the resulting plots where the boundaries of the 
theoretical $\beta$\,Cep (the hottest region in Fig.\,\ref{logTeff_logg}) and SPB instability 
strips have been taken from Miglio et al.\ (\cite{miglio}). 
A core convective overshooting parameter of 0.2 pressure scale heights was used in the stellar 
models since asteroseismic modeling results of $\beta$\,Cep targets have given evidence for the 
occurrence of core overshooting of that order (e.g. Aerts et al.\ \cite{aerts}). It is well-known 
that the choice of the metal mixture, opacities and metallicity also has a large influence on the 
extent of the instability regions. Here, we illustrate these domains for the OP (upper panel) 
and the OPAL (lower panel) opacity tables, as well as for two values of the metal mass fraction $Z$=0.01
(continuous boundaries of the instability regions) and $Z$=0.02 (dashed boundaries). 
However, we only adopt the metal mixture by Asplund et al. (\cite{asplund}) corrected with 
the Ne abundance determined by Cunha et al. (\cite{cunha}). Another choice of metal mixture 
(e.g., that by Grevesse \& Noels \cite{grevesse_noels}) leads to narrower instability domains, 
as shown in Miglio et al.\ (\cite{miglio}).

The stars in Fig.\,\ref{logTeff_logg} are marked by the letters given in Table\,\ref{positions}. The 
SB2 star KIC\,11\,973\,705 and the other presumed SB2 star KIC\,12\,207\,099 are marked by two 
asterisks, as their determined values are uncertain and no error bars can be given. It can be 
seen that 4 stars (b, e, j and p) fall into the middle of the SPB region. The two hottest 
stars (a and m) can be of SPB and/or $\beta$\,Cep nature. Both low order p- and g-modes, and 
high-order g-modes can be expected for these possibly 'hybrid' pulsators. Six of the stars 
(c, d, f, g, h and n) lie on or close to the boundaries of the SPB instability strips so 
that they possibly exhibit high-order g-mode pulsations. The remaining 4 stars lie outside 
the instability regions, two of them (k and i) are too cool to be main-sequence SPB stars 
and the two other ones (l and o) have too low \lg. Table\,\ref{positions} lists the potential pulsators 
together with their spectral types as derived in Sect.\,3.3. For the two suspected SB2 stars we
do not want to make a classification in terms of pulsators because their determined spectral types
may be completely wrong.

\section{Conclusions}
We tried to determine the fundamental parameters of B-type stars from the combined analysis of \hb\ and
the neighbouring metal lines in high-resolution spectra.  
Our results obtained for the test star Vega show that we can reproduce
the values of \te, \lg, \vs, metallicity and micro-turbulent velocity known from the literature and that our 
method works well at least in the 10\,000\,K range. 

The application of our programs to stars hotter than 15\,000\,K sets limitations in the accuracy of the results due to
the used LTE approximation. The independent analysis of the four hottest stars of our sample by NLTE-based
programs showed that the derived parameters agree within the errors of measurements for two of the stars, among
them the hottest star. The deviations obtained for the other two stars can be explained by other limitations of
the applied methods than the use of LTE and thus we believe that our results are valid within the derived
errors of measurement.  

In particular, the use of LTE cannot explain the large deviations in $T_{\rm eff}$ and 
$\log{g}$ following for some of the stars from the KIC data, however, as we showed on two examples. From our
results, there is strong evidence that the KIC systematically underestimates the temperatures of hotter stars, 
the difference increases with increasing \te. This finding confirms the results by Molenda-\.Zakowicz et al. 
(\cite{Molenda}) which
observed the same tendency for stars hotter than about 7\,000\,K.

The calculation of \te\ using SED-fitting based on the available photometric data revealed the reason
why the \te\ listed in the KIC are too low and why the difference is largest for the hottest stars: 
the stellar temperatures have been underestimated
because the interstellar reddening was not properly taken into account.

Eight stars of our sample show larger abundance anomalies.
Five of them have reduced metallicity, two are He-strong, one is He-weak,
and one is Si-strong.

According to our measurements, two of the 16 investigated stars fall into the overlapping range of 
the $\beta$\,Cep and SPB instability regions and could show, as so-called hybrid pulsators, both
low-order p- and g-modes and high-order g modes.
These are the two hottest stars in our sample, KIC\,3\,240\,411 and KIC\,10\,960\,750.
Four stars fall into the SPB instability region, and five more are located close to the borders 
of this region. 
The two coolest stars, KIC\,8\,451\,410, and KIC\,8\,583\,770, lie between
the SPB instability region and the blue edge of the classical instability strip. Two of the stars, 
KIC\,11\,973\,705 and KIC\,12\,207\,099, could not be classified because of their SB2 nature and one star, 
KIC\,8\,766\,405, is too evolved to show $\beta$\,Cep or SPB-type pulsations.

\begin{acknowledgements}
This research has made use of the SIMBAD database,
operated at CDS, Strasbourg, France, the Vienna Atomic Line Database (VALD),
and of data products from the Two Micron All
Sky Survey, which is a joint project of the University of Massachusetts
and the Infrared Processing and Analysis Center/California Institute of
Technology, funded by the National Aeronautics and Space Administration
and the National Science Foundation.
T.S. is deeply indebted to Dr. Yves Fr{\'e}mat for providing the NLTE specific intensity grids for this study.
A.T. and D.S. acknowledge the support of their work by the Deutsche Forschungsgemeinschaft (DFG), 
grants LE\,1102/2-1 and RE\,1664/7-1, respectively.

\end{acknowledgements}

\end{document}